\ifdefined\pdfminorversion\pdfminorversion=7\fi
\documentclass[11pt,a4paper]{article}

\usepackage[utf8]{inputenc}
\usepackage[a4paper,margin=2.5cm]{geometry}
\usepackage{amsmath,amssymb}
\usepackage{graphicx}
\IfFileExists{siunitx.sty}{\usepackage{siunitx}}{%
  \newcommand{\si}[1]{\ensuremath{\mathrm{##1}}}%
}
\usepackage{authblk}
\usepackage{subfig}
\usepackage{xcolor}
\usepackage[colorlinks=true,allcolors=blue]{hyperref}
\usepackage[capitalise,nameinlink]{cleveref}

\newcommand{\nsep}{n_{\mathrm{sep}}}
\newcommand{\rpol}{\rho_{\mathrm{pol}}}
\newcommand{\keywords}[1]{%
  \begin{center}\small\textbf{Keywords:} #1\end{center}%
}

\begin{document}

\title{Kinetic-Ballooning and Resistive X-Point Dynamics in Quasicontinuous Exhaust Tokamak Plasmas}

\author[]{Kaiyu Zhang\thanks{corresponding author: kaiyu.zhang@ipp.mpg.de}}
\author[]{Wladimir Zholobenko}
\author[]{Andreas Stegmeir}
\author[]{Michael Faitsch}
\author[]{Konrad Eder}
\author[]{Christoph Pitzal}
\author[]{Frank Jenko}
\author[]{ASDEX Upgrade Team\thanks{See appendix or author list of T. Pütterich et al., 2026 Nucl. Fusion 66 116002 https://doi.org/10.1088/1741-4326/ae61c8.}}
\affil{Max Planck Institute for Plasma Physics, Boltzmannstr. 2, 85748 Garching, Germany}
\date{}

\maketitle

\begin{abstract}
We investigate quasicoherent modes (QCMs) and their interaction with scrape-off-layer (SOL) plasma blobs (filaments) in the type-I-ELM-free quasicontinuous exhaust (QCE) regime, extending the analysis in [K. Zhang et al., Phys. Rev. Lett. 137, 055102 (2026)].
Two-fluid electromagnetic turbulence simulations are compared for two self-consistent turbulence states with different separatrix densities.
In both cases, a separatrix-spanning QCM resides within the radial-electric-field well, whereas blobs are observed only in the case with higher separatrix density.
The QCM exhibits similar kinetic-ballooning-mode (KBM) signatures in both cases, with modest differences in coherence and harmonic content.
The primary difference is the presence of an additional resistive X-point mode (RXM), which emerges as the near-X-point collisionality increases.
The coexistence of the KBM and RXM shifts the potential--density cross-phase from its KBM-like value toward the interchange value in the SOL, allowing QCM perturbations to be ejected as blobs.
These intermittent blobs account for over half of the net radial $E\times B$ transport in the far SOL.
Neither QCM onset nor blob ejection requires an increase in upstream collisionality.
The results distinguish the KBM-dominated mechanism that regulates the pedestal from the RXM-assisted blob-launching mechanism that enables intermittent SOL exhaust.
\end{abstract}

\keywords{tokamaks, quasicoherent mode, kinetic ballooning mode, resistive X-point mode, plasma blobs, electromagnetic turbulence}

\section{Introduction}

Reconciling high energy confinement with tolerable steady-state exhaust remains a central challenge for magnetic-confinement fusion~\cite{viezzer2023prospects}. Type-I edge-localized modes (ELMs) contribute to particle and impurity exhaust, but the associated transient heat loads pose a severe threat to plasma-facing components in reactor-scale devices such as ITER and DEMO~\cite{Loarte2003,eich2017elm}. Even in the absence of ELMs, the narrow scrape-off-layer (SOL) heat-flux footprint presents a major difficulty. Multi-machine regressions indicate that the SOL heat-flux decay length scales approximately as $\lambda_q\propto B_{\mathrm{pol}}^{-1}$~\cite{Eich2013}, leading to a millimeter-scale exhaust channel when extrapolated to ITER conditions~\cite{loarte2007power,Eich2013}. An attractive reactor regime must therefore combine an H-mode pedestal and the absence of large ELMs with sufficiently strong, continuous radial transport to broaden the exhaust footprint.

The enhanced-D$_\alpha$ (EDA) and quasicontinuous-exhaust (QCE) regimes are prominent examples of such type-I-ELM-free operation. EDA was first identified in Alcator C-Mod~\cite{Hubbard2001} and is characterized by a persistent electromagnetic fluctuation known as the quasicoherent mode (QCM). The QCM is localized around the pedestal foot, where it provides continuous particle transport, limits pedestal growth, and mitigates impurity accumulation. EDA is typically accessed at moderately high plasma shaping and separatrix density $\nsep$~\cite{gil2020stationary,gil2025eda}. 
QCE, first observed in ASDEX Upgrade (named type-II ELMs\cite{stober2001dependence} and later renamed to QCE~\cite{faitsch2021broadening}) and subsequently realized in TCV~\cite{labit2019dependence} and JET~\cite{faitsch2025quasi}, generally requires stronger shaping and higher $\nsep$~\cite{harrer2018parameter,faitsch2023analysis}. Its pedestal remains close to marginal stability against high-$n$ ballooning modes localized near the separatrix~\cite{radovanovicDevelopingPhysicsUnderstanding2022,harrerQuasicontinuousExhaustScenario2022}. QCE plasmas also exhibit QCM-like fluctuations, although their frequency spectra are generally broader than those measured in EDA~\cite{kalis2023experimental}. The experimentally observed continuous transition from EDA to QCE with increasing $\nsep$ suggests that a common QCM-like instability contributes to pedestal regulation in both regimes~\cite{Zohm2024,kalis2023experimental,schellpfeffer2026edge}.

Despite this apparent continuity, EDA and QCE differ substantially in their SOL dynamics. As $\nsep$ is increased toward QCE conditions, intermittent filamentary structures, hereafter referred to as blobs, become prominent in the SOL, accompanied by a substantial broadening of the heat-flux footprint~\cite{faitsch2021broadening,stagni2022dependence}. 
Experimental measurements for the ASDEX Upgrade (AUG) discharges yield radial filament velocities of order $1\,\si{km}/\si{s}$ in the QCE regime, comparable to those of type-I-ELM filaments and considerably larger than typical L-mode values\cite{grienerContinuousObservationFilaments2020}. 
The QCE blobs can therefore make a substantial contribution to radial particle and heat exhaust.
Comparable blob activity is weak or absent under EDA conditions~\cite{grienerContinuousObservationFilaments2020,faitsch2025quasi}. 
The coexistence of continuous pedestal transport and enhanced filamentary SOL transport makes QCE particularly attractive for power exhaust. At the same time, it raises two closely connected questions:

\paragraph{(1) Instability underlying the QCM.}

QCMs have been observed in both EDA and QCE regimes and studied extensively across several devices. Although the radial envelope of the QCM is generally understood to span the separatrix~\cite{kalis2023experimental,theiler2017radial}, measurements of its propagation direction and density--potential cross-phase $\alpha_{\phi,n}$ differ among devices, leading to different interpretations of the underlying instability.
Within a simplified classification based on linear instabilities, drift-wave-like fluctuations are characterized by $0<\alpha_{\phi,n}<\pi/2$ and typically propagate in the electron-diamagnetic direction. Kinetic-shear-Alfv\'en-wave-like fluctuations are characterized by $\pi/2<\alpha_{\phi,n}<\pi$ and are generally associated with propagation in the ion-diamagnetic direction. A cross-phase of $\alpha_{\phi,n}=\pi/2$ is characteristic of a pure interchange response, without a strong preference for either diamagnetic propagation direction.
In Alcator C-Mod, the QCM was initially interpreted as a resistive X-point mode~\cite{myra2005edge} and was subsequently associated with an electron drift wave with electromagnetic contributions~\cite{myra2002drift,snipes2001quasi,labombard2014new}. These interpretations were consistent with observations of propagation in the electron-diamagnetic direction and a small cross-phase, $\alpha_{\phi,n}\sim0$~\cite{labombard2014new}. Later measurements in Alcator C-Mod by Theiler \textit{et al.}~\cite{theiler2017radial}, however, indicated propagation in the ion-diamagnetic direction near the minimum of the radial electric field $E_r$, where the QCM is localized. Consistent with this result, measurements in AUG showed that the QCM propagates in the ion-diamagnetic direction within the $E_r$ well, with a plasma-frame phase velocity approximately one half of the ion diamagnetic velocity~\cite{kalis2023experimental}. The corresponding density--potential cross-phase is close to $\alpha_{\phi,n}\sim\pi$~\cite{kalis2023experimental,grenfell2024multi}. These signatures are more consistent with kinetic-shear-Alfv\'en dynamics and a kinetic ballooning mode (KBM) than with a conventional electron drift wave.
The diagnostic picture nevertheless changes outside the separatrix. In the AUG SOL, the measured cross-phase decreases toward $\alpha_{\phi,n}\sim0$~\cite{grenfell2024multi}, resembling the phase relation reported in some earlier Alcator C-Mod observations. The apparent differences among these measurements may therefore reflect a genuine radial variation in the QCM dynamics rather than mutually exclusive mode identifications. 

\paragraph{(2) Connection between the QCM and SOL blobs.}
The mechanism by which edge turbulence generates blobs depends on the confinement regime and remains incompletely understood~\cite{d2011convective}. Candidate mechanisms include nonlinear drift-wave dynamics~\cite{manz2015origin}, interchange instability~\cite{Furno2008}, and Kelvin--Helmholtz instability driven by shear in the $E\times B$ flow~\cite{gingellPlasmaBlobFormation2014}.
The QCM-identification and blob-launching problems are inseparable. 
If the same dominant instability drives QCM in both EDA-like and QCE states, what will enable QCM-scale perturbations to develop into radially propagating blobs in QCE but not under EDA conditions?
The onset of blobs must result either from a qualitative change in the nonlinear dynamics of QCM or from its interaction with the mean flow or an additional mode. 
These possibilities remain unresolved.
\\

We address these problems using global, nonlinear, full-$f$, two-fluid electromagnetic turbulence simulations performed with the \texttt{GRILLIX} code \cite{Stegmeir2019}.
Its locally field-aligned discretization maintains turbulence-scale resolution both across the separatrix and near the X-points~\cite{stegmeirFieldLineMap2016,stegmeirAdvancesFluxcoordinateIndependent2017,Stegmeir2023}.
By evolving the mean profiles and turbulent fluctuations self-consistently, these simulations capture both the mesoscopic pedestal motion associated with the QCM and the intermittent propagation of SOL blobs.
The main QCE simulation and its experimental validation were reported previously in a Letter~\cite{zhang2026turbulent}.
Here, we provide the detailed analysis underlying those results and extend the comparison to an EDA-like state in the same magnetic geometry.
This EDA-like state is obtained at a lower separatrix density and exhibits a QCM but no blobs.
The comparison helps clarify the physical nature of the QCM and the conditions for blob launching.
We further examine how collisionality and $E\times B$ shear regulate the QCM and blob dynamics and quantify the intermittent transport resulting from those blobs.


\section{Model and simulation setup}
The simulations are performed with \texttt{GRILLIX}, including magnetic-flutter~\cite{zhang2025dynamical}, a trans-collisional Landau-fluid heat-flux closure~\cite{pitzal2024low,Pitzal2023}, neoclassical flow and heat viscosities~\cite{zholobenko2024tokamak}, and a three-component neutral-gas model~\cite{eder2025self}. 
The code has been validated in attached L-mode on TCV \cite{Oliveira2022}
and across several AUG regimes: attached L-mode
\cite{zholobenkoRoleNeutralGas2021,zhang2024magnetic}, detached XPR L-mode
\cite{Eder2025X,Eder2026}, ITER baseline ELMy H-mode
\cite{zholobenko2024tokamak}, and ELM-free QCE \cite{zhang2026turbulent}.
It has also been used to study an L--H power ramp on AUG
\cite{Zholobenko2026}.
The magnetic equilibrium is taken from AUG discharge \#36165 at $t=6.1\,\si{s}$, during a QCE phase in a highly shaped, near-double-null configuration~\cite{faitsch2021broadening}. In the experiment, type-I ELMs disappear after $t=5\,\si{s}$ in a fuelling ramp while filament activity and the SOL temperature and power fall-off lengths increase.

\begin{figure}[!hbt]
    \centering
\subfloat[\label{fig:qce:omp_n}]{%
\begin{minipage}[c]{0.31\textwidth}
\centering
    \includegraphics[width=\textwidth]{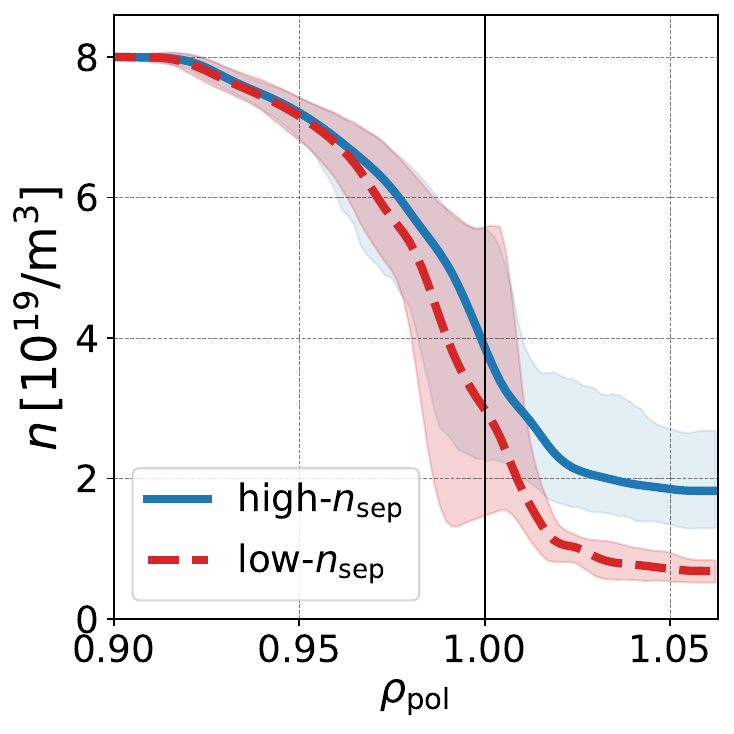}
    
\end{minipage}%
}
\subfloat[\label{fig:qce:omp_Te}]{%
\begin{minipage}[c]{0.31\textwidth}
\centering
    \includegraphics[width=\textwidth]{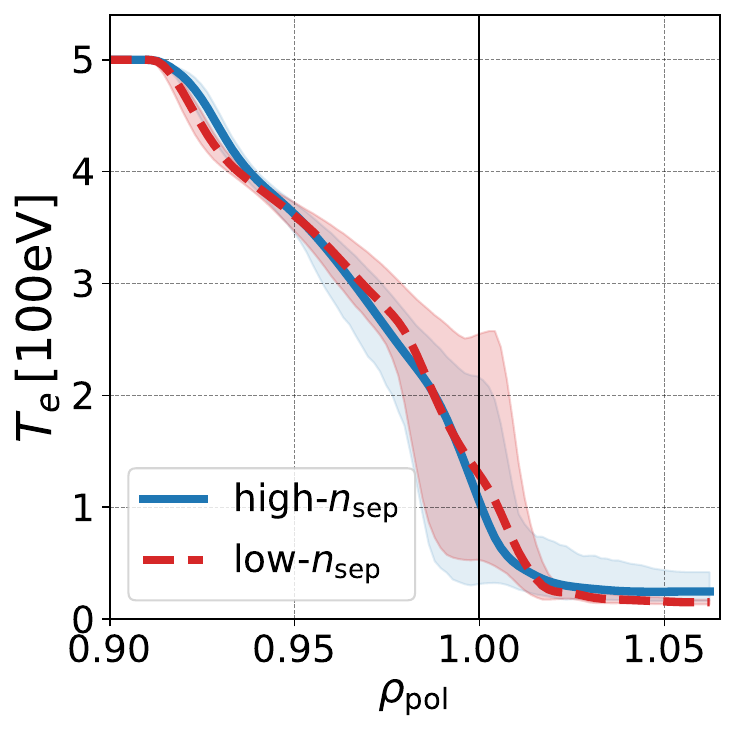}
    
\end{minipage}%
}
\subfloat[\label{fig:qce:omp_Ti}]{%
\begin{minipage}[c]{0.31\textwidth}
\centering
    \includegraphics[width=\textwidth]{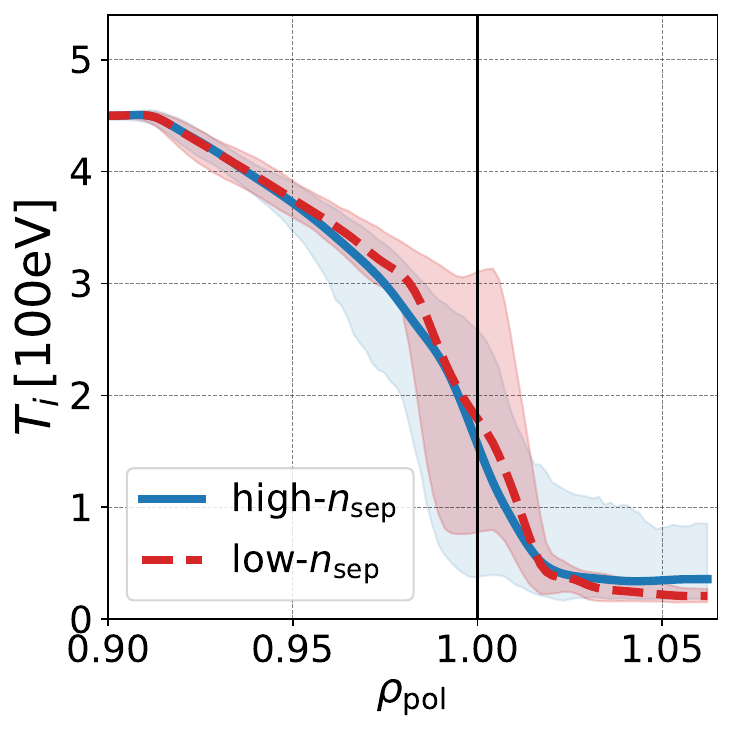}
    
\end{minipage}%
}
    \caption[Validation of profiles in QCE]{Density and temperature profiles at the outboard midplane. Solid lines represent the toroidally and temporally averaged values, while the shaded regions indicate the range of fluctuations, spanning the $1\%$ to $99\%$ quantiles.}
    \label{fig:qce:omp_profiles}
\end{figure}

We compare two simulations with identical magnetic geometry and core boundary conditions but different neutral boundary treatments. 
Figure~\ref{fig:qce:omp_profiles} compares the outboard-midplane profiles of $n$, $T_e$, and $T_i$. Solid curves are toroidal and temporal means; shaded bands span the 1st--99th percentiles. The adaptive source affects the region $\rpol<0.92$, whereas the separatrix and SOL profiles evolve freely.
In the first case, a particle-conserving plasma--neutral recycling model~\cite{eder2025self} yields the separatrix density $\nsep=4\times10^{19}/\si{m^{3}}$, together with strong fluctuations of QCM at the separatrix and broad SOL fluctuation envelopes. 
This case reproduces the measured mean profiles and fluctuation characteristics of discharge \#36165 at $t=6.1\,\si{s}$~\cite{zhang2026turbulent}, and is referred to as the high-$\nsep$ case in this study.
By contrast, in the low-$\nsep$ case, the neutral density is fixed at the divertor targets, following earlier \texttt{GRILLIX} studies~\cite{zholobenkoRoleNeutralGas2021,zhang2024magnetic,zholobenko2024tokamak}; the saturated density is $\nsep=3\times10^{19}\,/\si{m^{3}}$. 
It exhibits a stronger and more coherent QCM near the separatrix but no comparable blobs, only weak fluctuations farther into the SOL. 

In both simulations, the adaptive heat input to $0.9<\rpol<0.92$ saturates at approximately $3.2\,\si{MW}$ and maintains the core-boundary temperatures.
The global recycling fraction~\cite{zholobenkoRoleNeutralGas2021} is defined as $r_\mathrm{rcy}=1-S_n/(S_n+S_\mathrm{iz})$, where $S_n$ denotes the volume integral of the adaptive numerical particle source applied at the core boundary to maintain the prescribed density, and $S_\mathrm{iz}$ denotes the volume integral of the particle source due to neutral-gas ionization. 
In the saturated state, the recycling fractions are $r_\mathrm{rcy}=(99.95\pm0.1)\%$ and $r_\mathrm{rcy}=(99.8\pm0.2)\%$ for the high- and low-$\nsep$ cases, respectively. 
These values indicate that neutral-gas ionization supplies nearly all particles, with only a negligible contribution from the numerical source at the core boundary.
However, the global ionization rate, $S_\mathrm{iz}$, differs significantly between the two cases, reaching $1.3 \times 10^{23}/\si{s}$ in the high-$\nsep$ case, approximately 3.7 times the value of $3.5 \times 10^{22}/\si{s}$ in the low-$\nsep$ case.
We compute the key parameters relevant to the separatrix operational space~\cite{eich2021separatrix} in our simulations.
The separatrix density $n_{\mathrm{sep}}$ and electron temperature $T_{e, \mathrm{sep}}$ are obtained by time and toroidal averaging at the intersection of the separatrix and the outboard midplane.
The pressure fall-off length is computed as a flux-surface average, $\lambda_p = -\langle p \rangle/\langle \partial_r p \rangle$, evaluated at the separatrix and additionally averaged in time. 
Here, $p = nT_i + nT_e$ is the total plasma pressure, and $r$ denotes the minor radius.
The edge safety factor is $q_\mathrm{cyl} = 4.7$, and the effective ion charge is $Z_\mathrm{eff} = 1.24$. The major radius is $R_0 = 1.65\,\si{m}$. 
The $\lambda_p$-dependent MHD ballooning parameter follows the definition of Eq.(8) in Ref. \cite{faitsch2023analysis} and reads
\begin{equation}\label{eqn:alpha_mhd}
    \alpha_m = \frac{4\mu_0 R_0 q_{\mathrm{cyl}}^2 n_{\mathrm{sep}} T_{e,\mathrm{sep}}}{\lambda_p}\,.
\end{equation}
where $\mu_0$ is the vacuum permeability.
The collisionality follows the definition of Eq.(6) in Ref. \cite{faitsch2023analysis} and reads
\begin{equation}\label{eqn:nu}
    \nu^*_{e} = \frac{\pi q_{\mathrm{cyl}} R_0 Z_{\mathrm{eff}}}{1.03 \cdot 10^{16}} \frac{n}{T_{e}^{2}}\,.
\end{equation}
The two states have nearly the same pressure-gradient drive: their MHD ballooning parameters are $\alpha_m=2.3$, close to the equilibrium-dependent ideal-ballooning threshold $\alpha_c=2.7$ (computed by the linear MHD code \texttt{HELENA} \cite{radovanovicDevelopingPhysicsUnderstanding2022}). 
Their electron collisionalities differ much more strongly. 
At the outboard midplane (i.e. evaluating Eq.~\ref{eqn:nu} with $n_{\mathrm{sep}}$ and $T_{e,\mathrm{sep}}$), the low-$\nsep$ case has $\nu_{e,\mathrm{omp}}^*=5.2$, and the high-$\nsep$ case has $\nu_{e,\mathrm{omp}}^*=10$.
This pair of simulations is designed as a controlled comparison of two edge states, not as a scan in which $\nsep$ alone is varied. 
Changing the neutral boundary treatment modifies the ionization sources and, through the evolved profiles, $\nsep$, $T_{e,\mathrm{sep}}$, $T_{i,\mathrm{sep}}$, the gradient drive, collisionality, and the radial electric field, all of which can affect the microinstabilities. 
The time-dependent analysis in \cref{sec:qce:transition} will be used to determine which local parameters track the transitions in the QCM and blob dynamics.




\section{Edge dynamics across confinement regimes}\label{sec:qce:setup}
\begin{figure}
    \centering
\subfloat[\label{fig:qce:omp_n_tau_Lmode}]{%
\begin{minipage}{0.245\textwidth}
\centering
    \includegraphics[width=\textwidth]{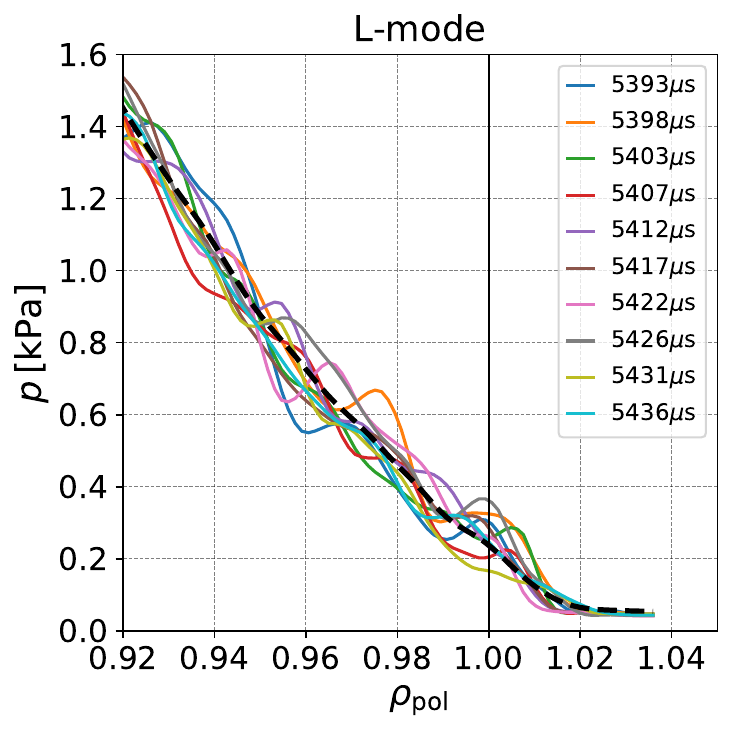}
    
\end{minipage}
}
\subfloat[\label{fig:qce:omp_n_tau_Hmode}]{%
\begin{minipage}{0.245\textwidth}
\centering
    \includegraphics[width=\textwidth]{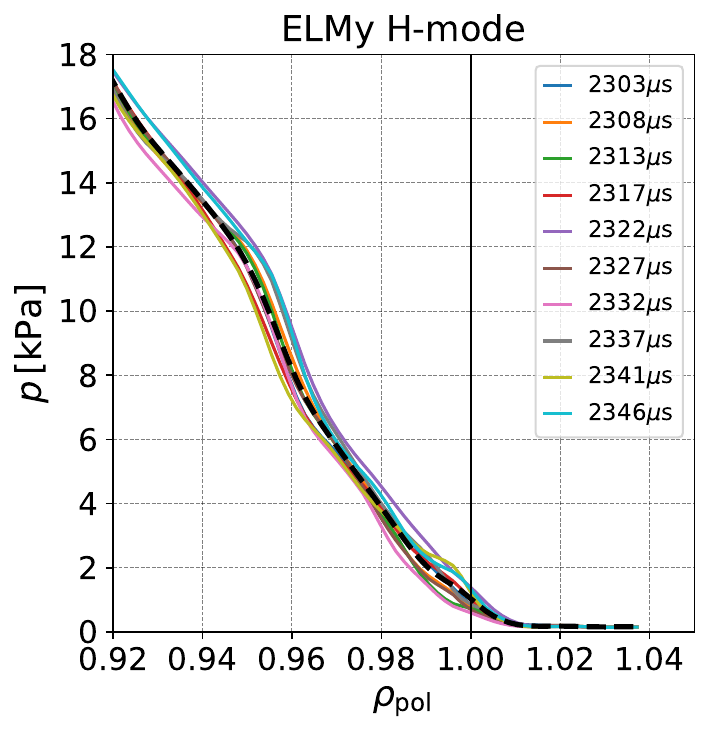}
    
\end{minipage}%
}
\subfloat[\label{fig:qce:omp_n_tau_EDA}]{%
\begin{minipage}{0.245\textwidth}
\centering
    \includegraphics[width=\textwidth]{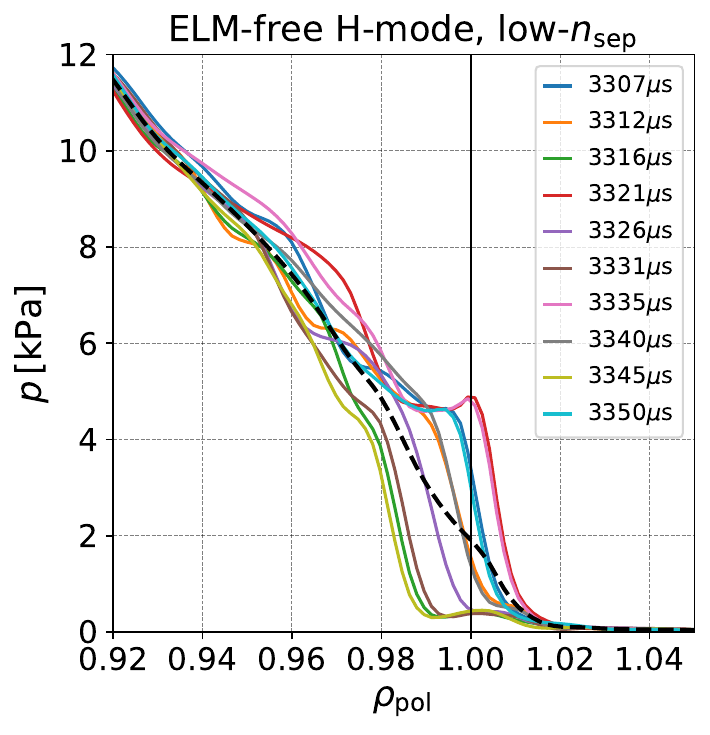}
\end{minipage}%
}
\subfloat[\label{fig:qce:omp_n_tau_QCE}]{%
\begin{minipage}{0.245\textwidth}
\centering
    \includegraphics[width=\textwidth]{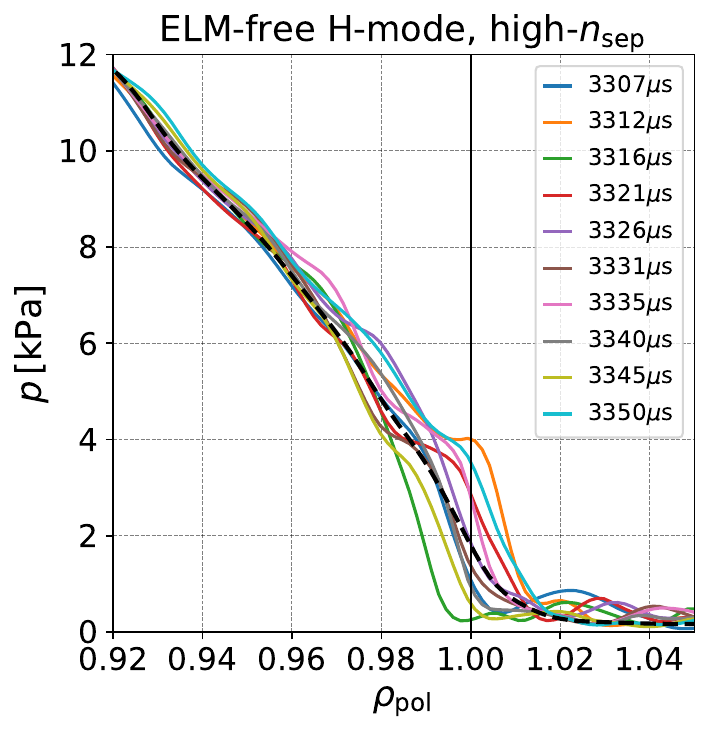}
\end{minipage}%
}
    \caption[Oscillating profiles across regimes]{Instantaneous outboard-midplane pressure profiles in (a) L-mode discharge \#38839, (b) inter-ELM H-mode discharge \#40411, (c) the low-$\nsep$ EDA-like case, and (d) the high-$\nsep$ QCE case. Successive curves are separated by approximately $5\,\si{\mu s}$.}
    \label{fig:qce:2D_pe_other}
\end{figure}

Both the low- and high-$\nsep$ cases use the same ELM-free magnetic equilibrium and exhibit a QCM. We first distinguish these two states from conventional L-mode~\cite{eder2025self} and inter-ELM H-mode~\cite{zholobenko2024tokamak} turbulence simulated with \texttt{GRILLIX}. 
Figure~\ref{fig:qce:2D_pe_other} shows instantaneous outboard-midplane profiles of the total pressure, $p=n(T_e+T_i)$, sampled approximately every $5\,\si{\mu s}$. 
In the L-mode case based on discharge \#38839~\cite{eder2025self}, broadband turbulence produces chaotic fluctuations throughout the edge and SOL. 
In the inter-ELM H-mode case based on discharge \#40411~\cite{zholobenko2024tokamak}, a steep pedestal begins near $\rpol=0.95$, and the instantaneous profiles remain tightly clustered around their mean, consistent with strong shear suppression of turbulence, a well-recognized feature of an H-mode pedestal.

\begin{figure}
    \centering
    \includegraphics[width=0.7\linewidth]{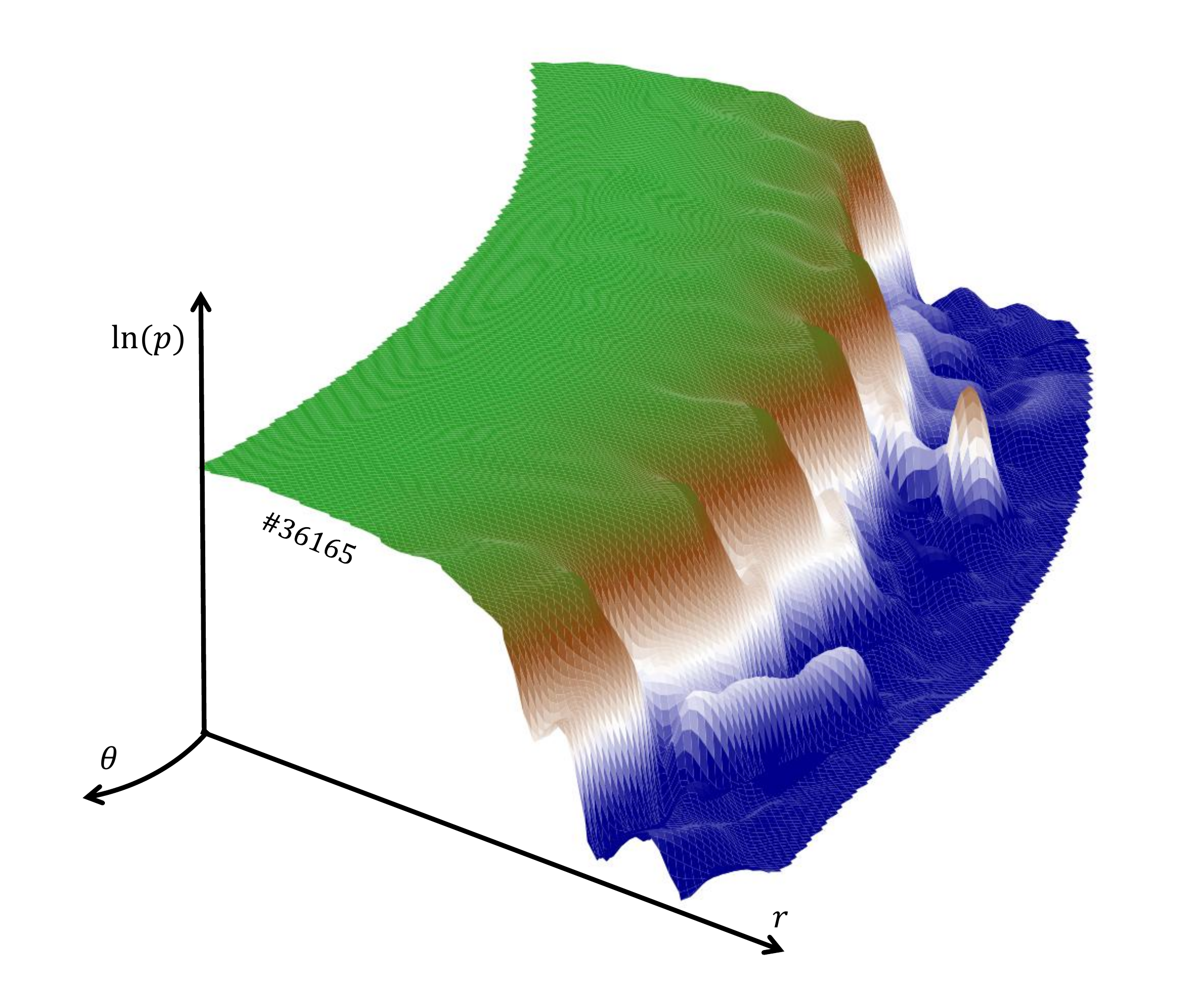}
    \caption[Mesoscopic deformation of the pedestal]{Topographic representation of the logarithmic plasma pressure on the low-field side in the high-$\nsep$ case. The QCM corrugates the pedestal foot, while detached structures propagate into the SOL.}
    \label{fig:qce:3d}
\end{figure}

The two ELM-free cases behave differently from the L-mode and inter-ELM H-mode scenarios. 
Their mean pedestals begin near $\rpol=0.97$, and the pedestal foot moves quasiperiodically across the separatrix while retaining a steep instantaneous gradient. 
In the high-$\nsep$ case, the envelope of instantaneous profiles extends farther into the SOL and contains outward-moving corrugations associated with blobs. 
The preserved instantaneous gradient shows that the QCM does not merely flatten the profile through local diffusion. 
Instead, it displaces the pedestal boundary over a mesoscopic radial distance comparable to the background gradient length.
The former can be represented by the radial autocorrection length $L_c=12 \si{mm}$ of QCM, while the latter is $\lambda_p=8 \si{mm}$, in the high-$\nsep$ case.
A fixed-position diagnostic therefore samples alternating high- and low-pressure states, producing the non-Gaussian distributions discussed in \cref{sec:qce:fila}. 
The three-dimensional representation in Figure~\ref{fig:qce:3d} makes the geometry of this motion explicit. 
The QCM corrugates the pressure pedestal while the radial gradient is preserved. 
As the mode propagates poloidally, an outboard-midplane measurement records the pedestal foot moving radially, as seen in Figure~\ref{fig:qce:2D_pe_other}. 

\section{Mode structures}\label{sec:qce:qcm}

\subsection{Field alignment and parity}
\begin{figure}
    \centering
\subfloat[\label{fig:qce:2D_align_pot}]{%
\begin{minipage}[b]{0.48\textwidth}
\centering
    \includegraphics[width=\textwidth]{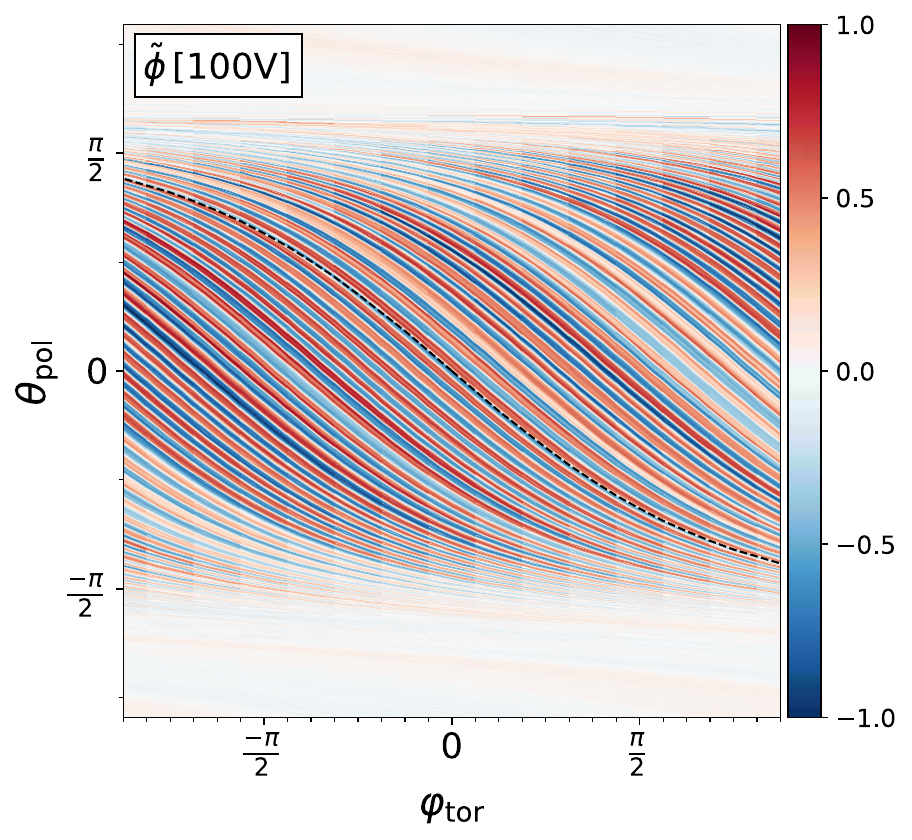}
    
\end{minipage}%
}
\subfloat[\label{fig:qce:2D_align_apar}]{%
\begin{minipage}[b]{0.48\textwidth}
\centering
    \includegraphics[width=\textwidth]{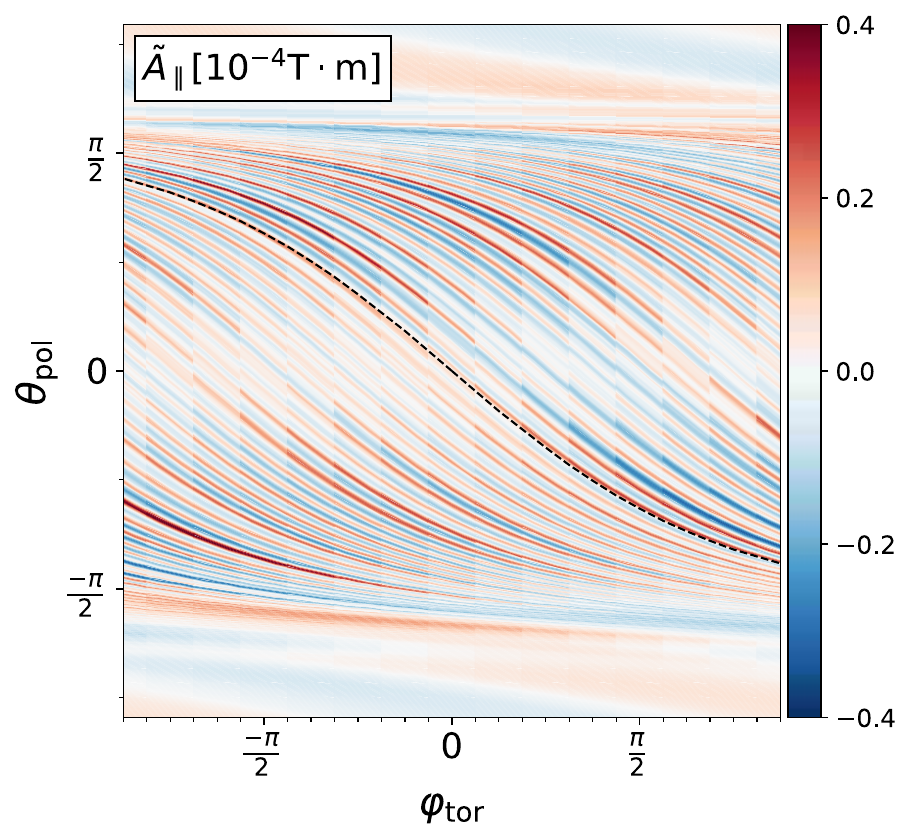}
    
\end{minipage}%
}
    \caption[Parity of the QCM]{Instantaneous fluctuations of (a) electrostatic potential $\tilde{\phi}$ and (b) parallel vector potential $\tilde{A}_\parallel$ on the unfolded surface $\rpol=0.999$ in the high-$\nsep$ case at $t=3.4\,\si{ms}$. Dashed lines follow the equilibrium magnetic field.}
    \label{fig:qce:aligned_999}
\end{figure}

Figure~\ref{fig:qce:aligned_999} shows the non-axisymmetric components $\tilde{\phi}=\phi-\langle\phi\rangle_\varphi$ and $\tilde{A}_\parallel=A_\parallel-\langle A_\parallel\rangle_\varphi$ on the unfolded surface $\rpol=0.999$ in the high-$\nsep$ case at $t=3.4\,\si{ms}$ (similar figures are also found in the low-$\nsep$ case). Here $\langle\cdot\rangle_\varphi$ denotes a toroidal average, and the outboard midplane (OMP) is at $\theta_\mathrm{pol}=0$. The calculation uses 16 toroidal planes, located at $\varphi_\mathrm{tor}=N\pi/16$ for $N=-8,\ldots,7$. Values between adjacent planes are obtained by nearest-plane continuation along field lines, without smoothing.
Both fields remain continuous along the dashed field-line trajectories, confirming that the mode is well aligned with the equilibrium magnetic field. The dominant pattern corresponds to a toroidal mode number $n=26$. With respect to the OMP, $\tilde{\phi}$ has even parity and $\tilde{A}_\parallel$ has odd parity. This parity combination is characteristic of twisting parity~\cite{ishizawa2019multi}. 
Field alignment and twisting parity establish the ballooning character of the fluctuation, but they do not by themselves distinguish an ideal, kinetic, or resistive ballooning branch. 

\subsection{Mode structure in Fourier space}
\label{sec:qce:qcm:fourier}
\begin{figure}[p]
    \centering
    \quad \quad
\subfloat[low-$\nsep$ \label{fig:qce:fourier_amp_EDA}]{%
\begin{minipage}[b]{0.465\textwidth}
\centering
    \includegraphics[width=\textwidth]{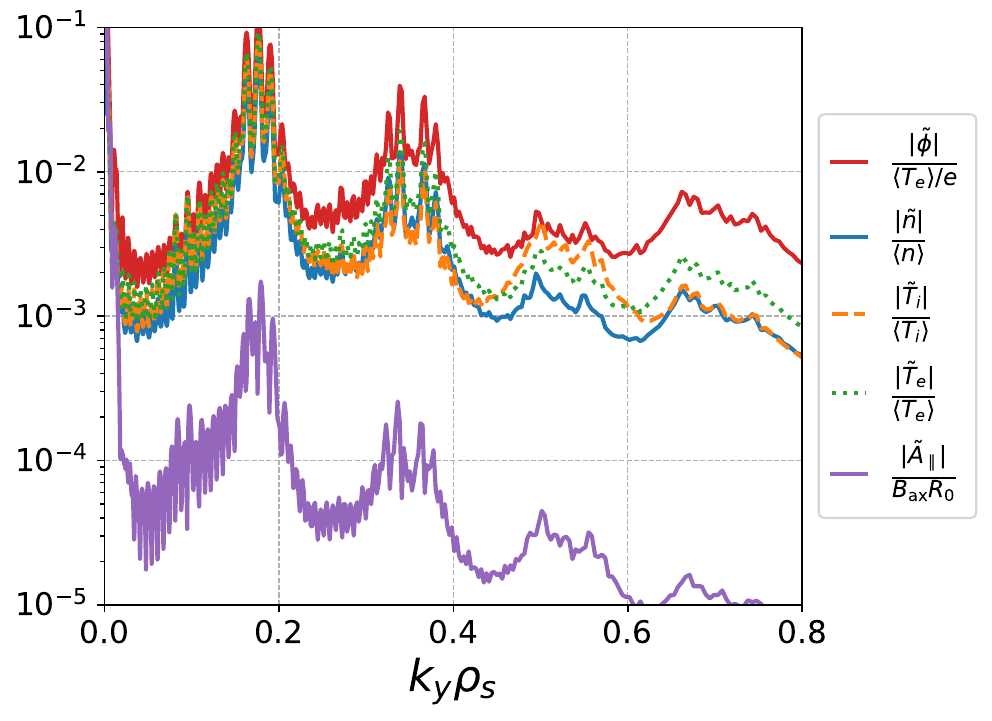}
    
\end{minipage}%
}
\subfloat[high-$\nsep$ \label{fig:qce:fourier_amp_QCE}]{%
\begin{minipage}[b]{0.465\textwidth}
\centering
    \includegraphics[width=\textwidth]{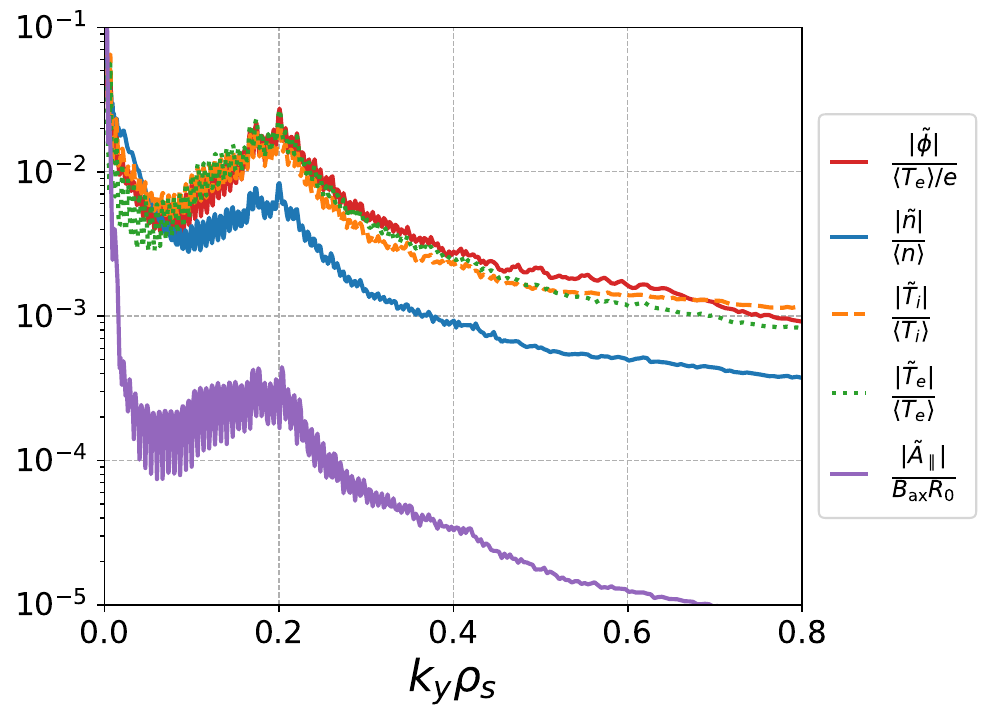}
    
\end{minipage}%
}

\subfloat[low-$\nsep$ \label{fig:qce:fourier_shift_EDA}]{%

\begin{minipage}[b]{0.46\textwidth}

\centering
    \includegraphics[width=\textwidth]{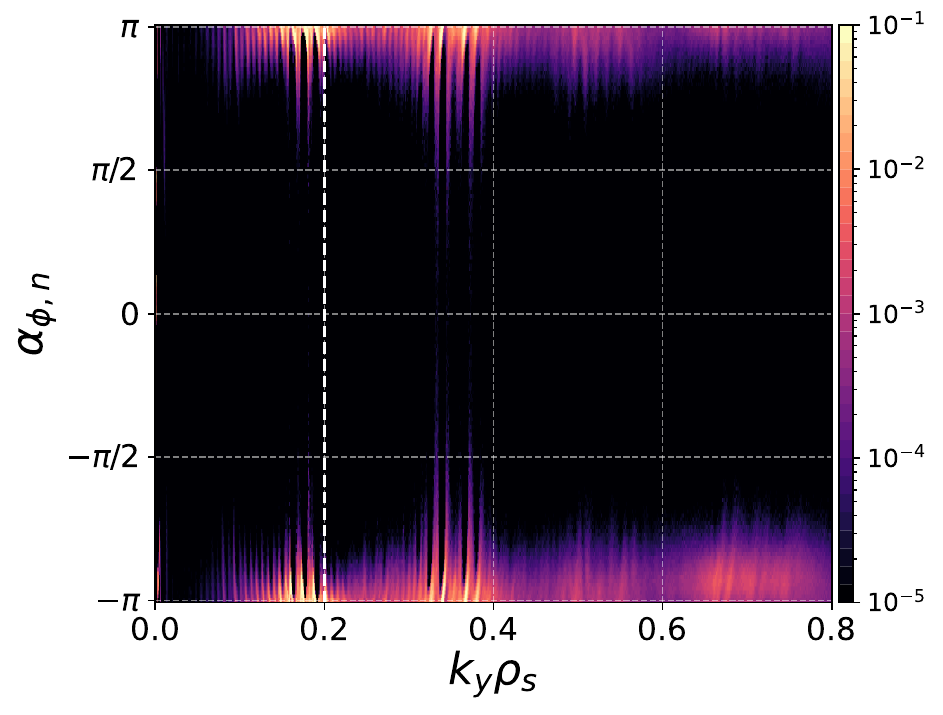}

\end{minipage}%

}
\subfloat[high-$\nsep$ \label{fig:qce:fourier_shift_QCE}]{%
\begin{minipage}[b]{0.46\textwidth}
\centering
    \includegraphics[width=\textwidth]{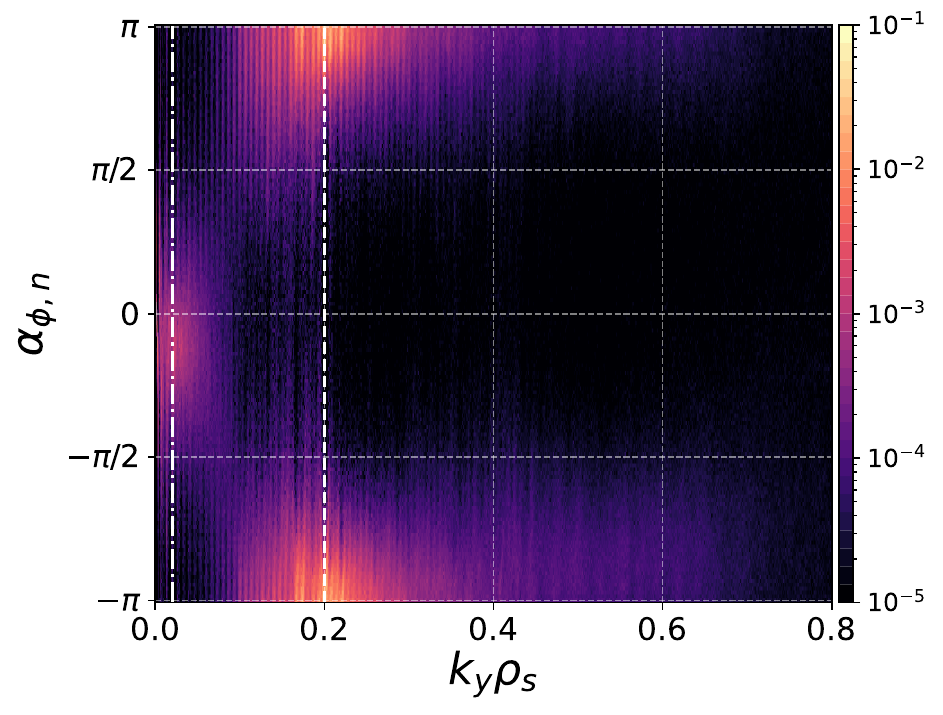}
    
\end{minipage}%
}

\subfloat[low-$\nsep$ \label{fig:qce:fourier_disp_EDA}]{%

\begin{minipage}[b]{0.46\textwidth}

\centering
    \includegraphics[width=\textwidth]{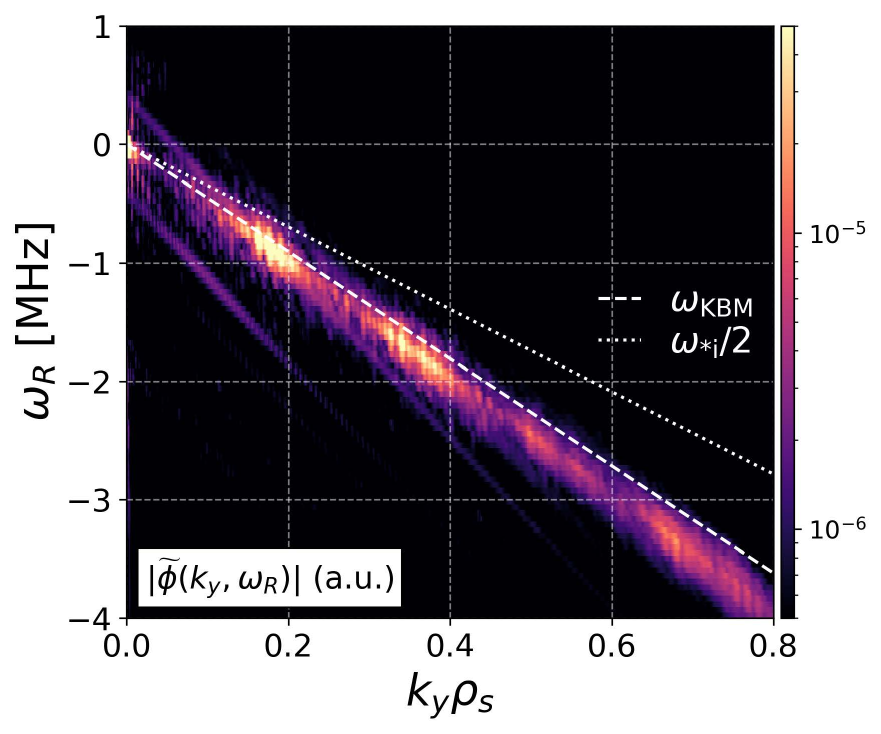}

\end{minipage}%

}
\subfloat[high-$\nsep$ \label{fig:qce:fourier_disp_QCE}]{%
\begin{minipage}[b]{0.46\textwidth}
\centering
    \includegraphics[width=\textwidth]{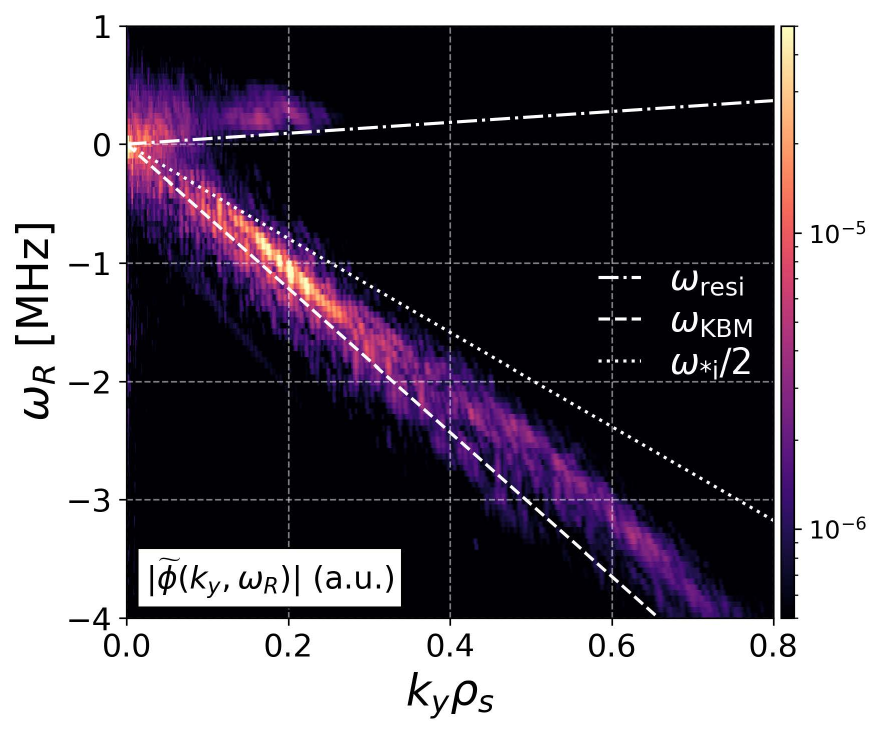}
    
\end{minipage}%
}
\caption[Fourier spectra of QCM]{Binormal Fourier spectra at the flux surface $\rpol=0.999$, comparing the low-$\nsep$ case (left column) and the high-$\nsep$ case (right column).  
Panels (a) and (b): Fourier amplitudes of selected fluctuating quantities.  
Panels (c) and (d): histograms of the phase shift between $\tilde{n}$ and $\tilde{\phi}$.  
Panels (e) and (f): dispersion relations in the plasma frame, represented by the amplitudes of $\tilde{\phi}$.  
The dashed line and dash-dot line indicate the kinetic-ballooning-mode (KBM) dispersion and the resistive mode dispersion, respectively, evaluated from the local equilibrium and profile parameters.}\label{fig:qce:fourier_amp}
\end{figure}

Figure~\ref{fig:qce:fourier_amp} compares binormal spectra at $\rpol=0.999$. The left and right columns show the low- and high-$\nsep$ cases, respectively; the rows give fluctuation amplitudes, potential--density cross-phases, and plasma-frame dispersion relations. 
Because the mode is field-aligned, the transform is performed in the symmetry-flux coordinate $\theta_y$~\cite{ribeiro2010conformal}. 
This angle straightens magnetic-field lines in the toroidal direction, and $k_y$ denotes its conjugate binormal wavenumber. 
It is worth noticing that experimental measurements generally report the conventional OMP poloidal wavenumber $k_\mathrm{pol}$~\cite{kalis2023experimental}. 
The two scales are related by $k_\mathrm{pol}/{k_y}={\mathrm{d}\theta_y}/{\mathrm{d}\theta_\mathrm{pol}}$, which is approximately $1/6.05$ at the outboard midplane for this magnetic equilibrium.
The transformation between the symmetry coordinate $\theta_y$ and the geometric poloidal angle $\theta_\mathrm{pol}$ is shown in Figure~\ref{fig:qce:theta_y_pol} in the Appendix.

\subsubsection{Common ground: Kinetic Ballooning Mode}
In both cases, the spectra of $\phi$, $n$, $T_e$, $T_i$, and $A_\parallel$ peak at $k_y\rho_s=0.2$ [Figures~\ref{fig:qce:fourier_amp_EDA} and \ref{fig:qce:fourier_amp_QCE}]. The finite $A_\parallel$ amplitude confirms that the QCM is electromagnetic. 
Under the OMP coordinate conversion in Figure~\ref{fig:qce:theta_y_pol}, the peak becomes $k_\mathrm{pol}\rho_s=0.033$, within the measured AUG range $0.025$--$0.075$~\cite{kalis2023experimental}.
The potential--density cross-phase,
$\alpha_{\phi,n}=\arg(\tilde{\phi}_{k_y})-\arg(\tilde{n}_{k_y})$, is close to $\pi$ at the QCM scale in both cases. 
The temperature cross-phases satisfy $\alpha_{\phi,T_e}\approx\alpha_{\phi,T_i}\approx\pi$, so density and temperature perturbations are approximately in phase. This behavior is characteristic of kinetic-shear-Alfv\'en dynamics and of a marginal KBM when finite-Larmor-radius polarization is retained~\cite{scott2021turbulence,doerk2016gyrokinetic}; it is also consistent with recent probe measurements of QCM inside the separatrix in AUG~\cite{grenfell2024multi}.
By contrast, the cross-phase $\alpha_{\phi,n}=\pi/2$, characteristic of an ideal interchange response, would maximize the radial $E\times B$ flux at fixed amplitudes.
The plasma-frame frequency is obtained by subtracting the flux-surface-averaged $E\times B$ advection,
$
    \omega_r = \omega_\mathrm{lab}
    -k_y\frac{1}{2\pi}\int_0^{2\pi}v_E^y\,\mathrm{d}y\,.
$
Negative $\omega_r$ denotes ion-diamagnetic propagation.
Figure~\ref{fig:qce:fourier_disp_QCE} shows the frequency spectrum in the high-$\nsep$ case, corresponding to the result reported in Fig.~3 of Ref.~\cite{zhang2026turbulent}; Figure~\ref{fig:qce:fourier_disp_EDA} provides the comparison with the low-$\nsep$ case. 
In both cases, the dominant branch is negative and follows the KBM estimate shown by the dashed curves in Figures~\ref{fig:qce:fourier_disp_EDA} and \ref{fig:qce:fourier_disp_QCE}.
Two common instabilities propagate in the ion-diamagnetic direction: the ion-temperature-gradient mode (ITG) and the kinetic ballooning mode (KBM).
For the two-fluid model used here, the long-wavelength asymptotic dispersion yields $\omega_\mathrm{itg}\sim \sqrt{\omega_B\omega_{i*}}$ for the ITG mode and $\omega_\mathrm{kbm}\sim \omega_{i*}/2$ for the KBM \cite{aleynikova2017quantitative,zhang2026electromagnetic}.
Here $\omega_{i*}$ is the ion diamagnetic frequency and $\omega_B$ is the magnetic drift frequency. 
For the edge parameters of the present simulations, $|\omega_\mathrm{itg}/\omega_\mathrm{kbm}|\sim|\omega_B/\omega_{i*}|^{1/2}\sim0.1$. 
The two frequencies are therefore well separated, making an ITG interpretation unlikely. 
Together with the twisting parity, electromagnetic amplitude, and $\alpha_{\phi,n}\sim \pi$, these results identify the KBM as the dominant driver of the QCM.
\begin{figure}
    \centering
\subfloat[low-$\nsep$\label{fig:qce:eigen_fluct_theta_l}]{%
\begin{minipage}[b]{0.465\textwidth}
\centering
    \includegraphics[width=\textwidth]{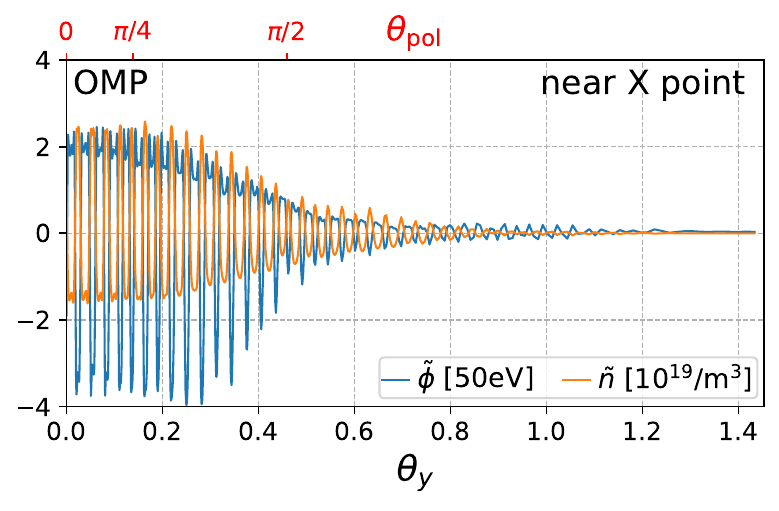}
    
\end{minipage}%
}
\subfloat[high-$\nsep$\label{fig:qce:eigen_fluct_theta_h}]{%
\begin{minipage}[b]{0.465\textwidth}
\centering
    \includegraphics[width=\textwidth]{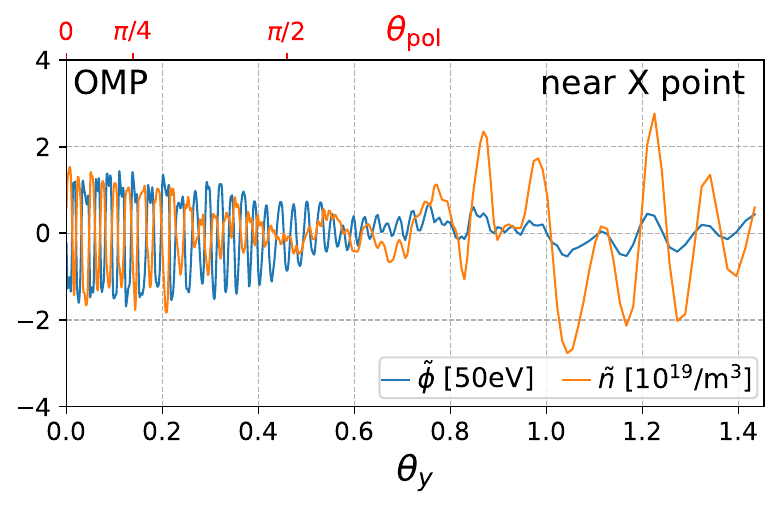}
    
\end{minipage}%
}
\caption[Identification of X-point fluctuations]{Potential fluctuations $\tilde{\phi}$ and density fluctuations $\tilde{n}$ along the $\rpol=0.999$ flux surface at $t=3.41\,\si{ms}$ for (a) the low-$\nsep$ case and (b) the high-$\nsep$ case.
The bottom and top axes show the symmetry-flux angle $\theta_y$ and the geometric poloidal angle $\theta_\mathrm{pol}$, respectively.
The plotted interval extends from the outboard midplane ($\theta_y=\theta_\mathrm{pol}=0$) to the point of minimum $Z$ on the flux surface near the X point.}
\label{fig:qce:eigen_fluct_theta}
\end{figure}
\subsubsection{Distinction: Resistive X-point mode}
As shown in Figures~\ref{fig:qce:fourier_amp_EDA} and \ref{fig:qce:fourier_amp_QCE}, the high-$\nsep$ spectra are broader around $k_y\rho_s=0.2$ and extend farther along the KBM dispersion curve. 
The low-$\nsep$ ones are more coherent and contain clear harmonics at $k_y\rho_s\approx0.18$, $k_y\rho_s\approx0.18\times 2$, $k_y\rho_s\approx0.18\times 3$, and $k_y\rho_s\approx0.18\times 4$.
The frequencies of those higher harmonics are also equally distant, as shown in Figure~\ref{fig:qce:fourier_disp_EDA}.

More importantly, only the high-$\nsep$ case contains a secondary branch in Figure~\ref{fig:qce:fourier_disp_QCE}, which propagates slowly in the electron diamagnetic direction and was interpreted as a resistive X-point mode (RXM) in Ref.~\cite{zhang2026turbulent}. 
Several pieces of evidence support this RXM interpretation. 
\paragraph{Frequency} The small real frequency observed in Figure~\ref{fig:qce:fourier_disp_QCE} is consistent with the resistive-mode estimate
$\omega_\mathrm{resi}\sim ({\omega_{e*}+\omega_{i*}})/{3}\,$ \cite{morales2016edge}.
The sign of $\omega_\mathrm{resi}$ is not essential because it depends on the balance between the electron and ion pressure-gradient contributions. Since $\omega_{e*}>0$ and $\omega_{i*}<0$, their partial cancellation keeps the real frequency close to zero and gives no strong directional preference. 
Consequently, this resistive branch remains spectrally distinct from the KBM, with $|\omega_\mathrm{resi}|\ll|\omega_\mathrm{kbm}|$.

\paragraph{Wavenumber} Further evidence arises from the wavenumber characteristics. 
The resistive ballooning mode (RBM) has a characteristic wavelength $ L_\mathrm{rbm}=2\pi q R/\sqrt{1.96\tau_e\omega_{ci}\omega_{ce}t_{rbm}}$ ~\cite{zeiler2002tokamak}, where $\tau_e$ is the collision time, $\omega_{ci}$ and $\omega_{ce}$ are ion and electron cyclotron frequencies, respectively. 
And $t_{rbm}=c_s^{-1}\sqrt{R_0L_n/2}$ is the characteristic time scale of RBM.
RBMs will only grow at shorter wavelengths than $L_\mathrm{rbm}$~\cite{zeiler2002tokamak}.
In contrast, a resistive mode with longer wavelength will no longer be restricted in the bad curvature region, and therefore its ballooning characteristic will be lost~\cite{zeiler2002tokamak}.
In our high-$\nsep$ simulation, the resistive mode appears at $k_y\rho_s=0.02$--$0.2$, corresponding to the wavelength $L\approx31\rho_s$--$ 310\rho_s$.
These values are much longer than the wavelength $L_\mathrm{rbm}\approx14\rho_s$ expected for the unstable RBM.
This suggests this mode is more likely a resistive mode near the X-point region than an RBM that peaks at the outboard midplane.
\paragraph{Phase shift} The RXM satisfies $-\pi/4<\alpha_{\phi,n}<\pi/4$ and therefore has a drift-wave-like phase relation, distinct from the interchange-like value $\alpha_{\phi,n}\sim\pi/2$ expected for an RBM. 
This can be understood by the effectiveness of the interchange force in setting the phase shift.
The phase shift of the interchange mode is largely set by the ion polarization and the curvature of pressure in the vorticity equation, which reads $\nabla_\perp^2\tilde{\phi}\propto -\mathcal{C}(\tilde{n})$.
RBM peaks at the outward plane, where $\nabla_\perp^2 \sim -k_y^2$ and $\mathcal{C}\sim i k_y$.
Hence, after canceling $-k_y$ on both sides, we have $k_y \tilde{\phi}\propto i\tilde{n}$, resulting a cross-phase around $\pi/2$.
However, for RXM around the X-point, $\nabla_\perp^2\propto -k_y^2 $ while $\mathcal{C}\sim i k_x$.
This leads to $k_y^2\tilde{\phi}\propto ik_x\tilde{n}$.
Notice that $ik_x\tilde{n}$ will not change the cross phase defined in the $y$ direction. Therefore, we have $\alpha_{\phi,n}\sim 0$.

\paragraph{Spatial origin}
Figure~\ref{fig:qce:eigen_fluct_theta} shows the eigenfunction-like fluctuation structures on the $\rpol=0.999$ flux surface, which we use to identify where the modes are localized. The bottom axis shows the symmetry-flux angle $\theta_y$, the coordinate used to define the Fourier wavenumbers in Section~\ref{sec:qce:qcm:fourier}. The top axis marks the corresponding poloidal angle $\theta_\mathrm{pol}$, with $\theta_y=\theta_\mathrm{pol}=0$ at the outboard midplane. The right edge of each panel, at $\theta_y\approx1.5$, corresponds to the point of minimum $Z$ on this flux surface.
The mapping between the two angles is nonlinear, as shown in Figure~\ref{fig:qce:theta_y_pol} in the Appendix. The region near the X point ($\theta_\mathrm{pol}>\pi/2$) spans a wide interval in $\theta_y$. Here, $\mathrm{d}\theta_\mathrm{pol}/\mathrm{d}\theta_y\approx0$, so even a mode with a short poloidal wavelength can have a long wavelength in $\theta_y$ and thus appear at $k_y\approx0$.
In both the low- and high-$\nsep$ cases (Figures~\ref{fig:qce:eigen_fluct_theta_l} and~\ref{fig:qce:eigen_fluct_theta_h}, respectively), the KBM fluctuations peak at the outboard midplane and exhibit an anticorrelation between $\tilde{\phi}$ and $\tilde{n}$. A secondary mode localized near the X point appears only in the high-$\nsep$ case. The positive correlation between $\tilde{\phi}$ and $\tilde{n}$ in this region is consistent with the near-zero cross-phase ($\alpha_{\phi,n}\approx0$) at small $k_y$ in Figure~\ref{fig:qce:fourier_shift_QCE} and a secondary branch propagating in the electron direction ($\omega_r>0$) in Figure~\ref{fig:qce:fourier_disp_QCE}. Together, the spatial localization, cross-phase, and propagation direction support the identification of this secondary mode as an RXM. 
In the low-$\nsep$ case, the KBM fluctuations gradually weaken toward the X point, while $\tilde{\phi}$ and $\tilde{n}$ remain anticorrelated throughout the plotted region.


No single observable uniquely identifies the mode. 
The RXM inference is instead based on the conjunction of its frequency, wavenumber, phase relation, and spatial localization as demonstrated in Figures~\ref{fig:qce:fourier_amp} and \ref{fig:qce:eigen_fluct_theta}. 
The KBM is driven primarily in the bad-curvature OMP region, whereas the resistive branch grows where $k_\perp$ and collisional magnetic diffusion are amplified near the X-point. 
Their separation explains how both branches can coexist on the same flux surface without the RXM replacing KBM, or vice versa.
The shaping dependence of the QCE regime, in particular its larger triangularity and safety factor, increases the connection length from the high-field side to the low-field side, $l_{\mathrm{HFS} \rightarrow \mathrm{LFS}}$~\cite{harrerQuasicontinuousExhaustScenario2022}.
This may also facilitate the separation of X-point modes from OMP modes inside the separatrix.

    
    

\section{Blob dynamics}\label{sec:qce:fila}
\subsection{Identification}

\begin{figure}
    \centering
    \includegraphics[width=1\linewidth]{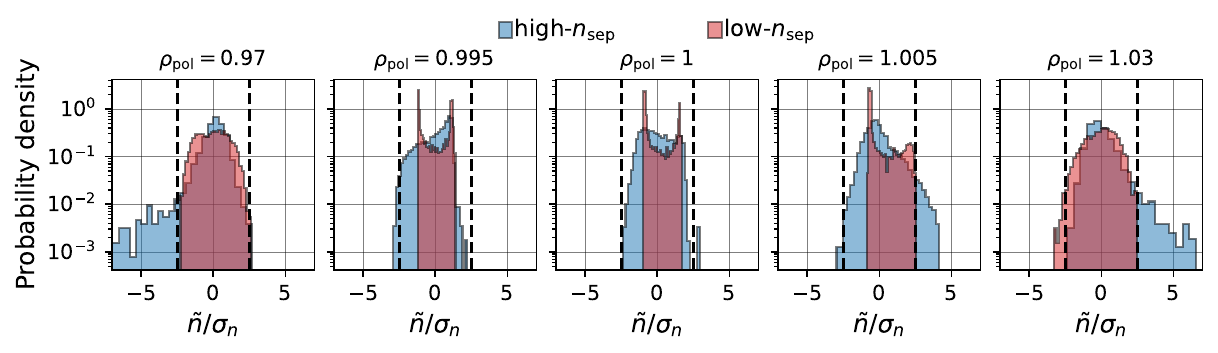}
    \caption[Radial evolution of density-fluctuation statistics]{Probability densities of the normalized OMP density fluctuation $\tilde n/\sigma_n$ at selected radii. Blue and red distributions correspond to the high- and low-$\nsep$ cases. Vertical dashed lines mark $\tilde n/\sigma_n=\pm2.5$, the thresholds used to classify voids and blobs.}
    \label{fig:qce:pdf_rho_list}
\end{figure}

Large-amplitude, field-aligned filaments---commonly called blobs---are coherent SOL structures with perpendicular sizes of centimeters and parallel extents of meters~\cite{d2011convective}. 
A major criterion for identifying blobs is intermittency.
At every spatial point, the fluctuation and its local standard deviation are defined by
\begin{equation}
    \tilde n=n-\langle n\rangle_{t,\varphi},
    \qquad
    \sigma_n=\left\langle\tilde n^2\right\rangle_{t,\varphi}^{1/2},
\end{equation}
where $\langle\cdot\rangle_{t,\varphi}$ denotes a temporal and toroidal average, and $\tilde n/\sigma_n$ is the intermittency variable. 
Blob samples satisfy $\tilde{n}/\sigma_n>2.5$. Samples with $\tilde{n}/\sigma_n<-2.5$ are classified as voids, and $-2.5<\tilde{n}/\sigma_n<2.5$ defines the non-intermittent background.
Figure~\ref{fig:qce:pdf_rho_list} compares the probability density of $\tilde{n}/\sigma_n$ at five OMP radii. 
At $\rpol=0.995$, near the QCM, both cases are non-Gaussian. 
In particular, the low-$\nsep$ distribution is distinctly bimodal because the oscillating pedestal repeatedly carries strong fluctuations across a fixed radial location. 
The local signal therefore alternates between pedestal-top and SOL values, as shown in Figure~\ref{fig:qce:omp_n_tau_EDA}.
Away from the QCM, the two cases diverge clearly. 
In the low-$\nsep$ case, the distributions are approximately symmetric and remain largely within $|\tilde{n}/\sigma_n|<2.5$. 
In the high-$\nsep$ case, the confined-side distribution at $\rpol=0.97$ develops a long negative tail, whereas the SOL distributions at $\rpol=1.005$ and $1.03$ develop long positive tails. 
The former identifies voids and the latter blobs. 
The simultaneous appearance of inward voids and outward blobs is consistent with pressure-gradient relaxation producing paired perturbations~\cite{cao2025physics}.
We therefore identify blobs only in the high-$\nsep$ case, not in the low-$\nsep$ case. 

\subsection{Transport quantification}

\begin{figure}[!ht]
    \centering
    \includegraphics[width=0.7\linewidth]{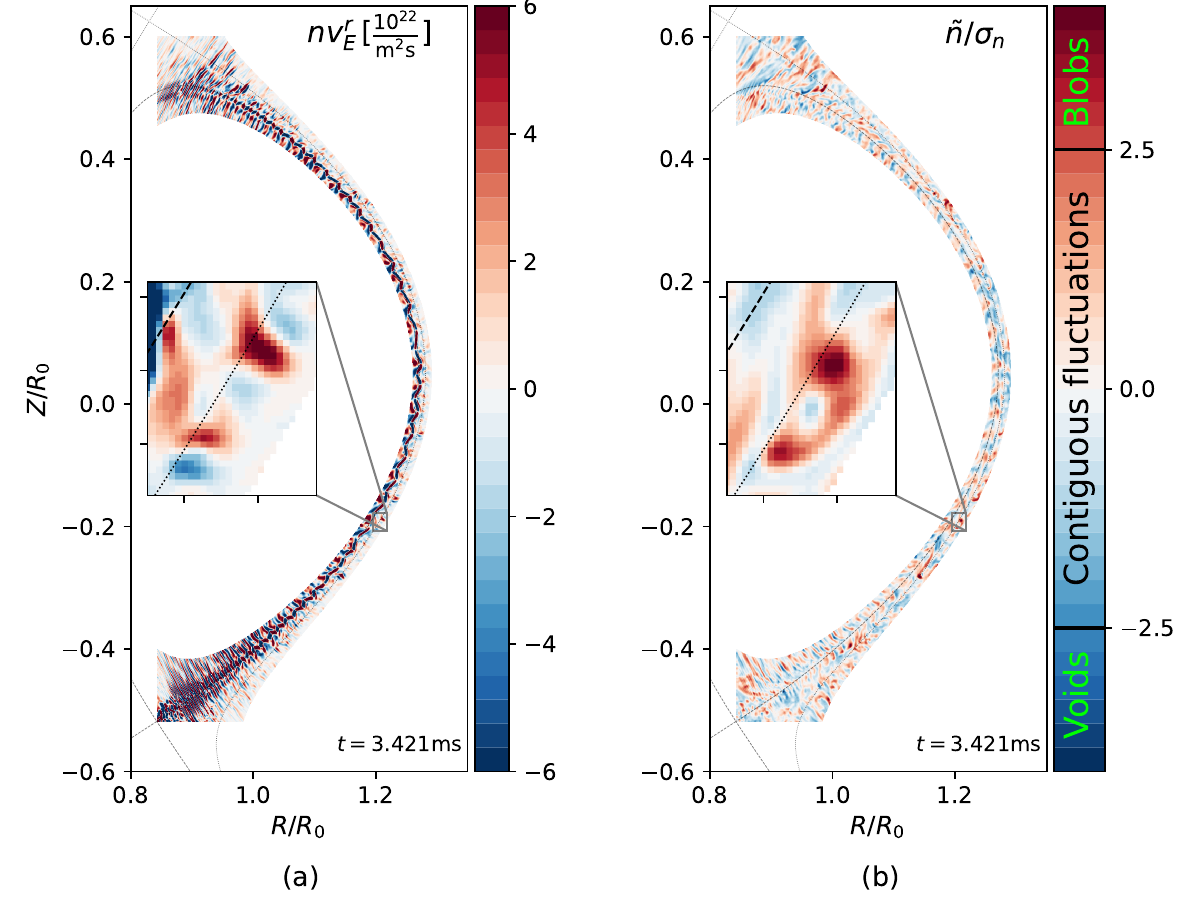}
\caption[Spatial distribution of intermittent transport]{Low-field-side snapshot of (a) the radial $E\times B$ particle flux $n v_E^r$ and (b) the normalized density fluctuation $\tilde n/\sigma_n$ in the high-$\nsep$ case. The black dashed and dotted curves denote the primary and secondary separatrices, respectively.}
    \label{fig:qce:2D_gamma_exb_n_2plot}
\end{figure}

We now quantify how much of the radial particle flux is carried by the blobs identified above. The analysis is restricted to the low-field-side domain in Figure~\ref{fig:qce:2D_gamma_exb_n_2plot}, where the dominant fluctuations occur and flux surfaces can be compared consistently before the magnetic topology divides beyond the secondary separatrix. The diamagnetic particle flux vanishes at the separatrix, while magnetic-flutter particle transport is confined to a narrow cross-separatrix layer and is similar in both cases. We therefore focus on the radial $E\times B$ flux $n v_E^r$.
Figure~\ref{fig:qce:2D_gamma_exb_n_2plot} compares $n v_E^r$ and $\tilde{n}/\sigma_n$ in the high-$\nsep$ case at $t=3.42\,\si{ms}$. The near-separatrix flux is large but mostly associated with $|\tilde{n}/\sigma_n|<2.5$. Strongly intermittent positive events occur farther into the SOL.
In the zoomed inset of Figure~\ref{fig:qce:2D_gamma_exb_n_2plot}(b), we identify two blobs.
The upper one coincides with a coherent region of positive $n v_E^r$, directly demonstrating outward convection. 
The lower event instead has a dipolar flux pattern consistent with a spinning structure. 
Finite-$T_e$ sheath coupling can align $\tilde\phi$ with $\tilde T_e$ and close the potential contours around a hot blob, producing such rotation~\cite{d2011convective}.
Thus, not every blob contributes the same instantaneous radial flux; a statistical decomposition is required.

\begin{figure}[!tb]
    \centering
\subfloat[low-$\nsep$ \label{fig:qce:intermitt_exb_rho_l}]{%
\begin{minipage}[c]{0.46\textwidth}
\centering
    \includegraphics[width=\textwidth]{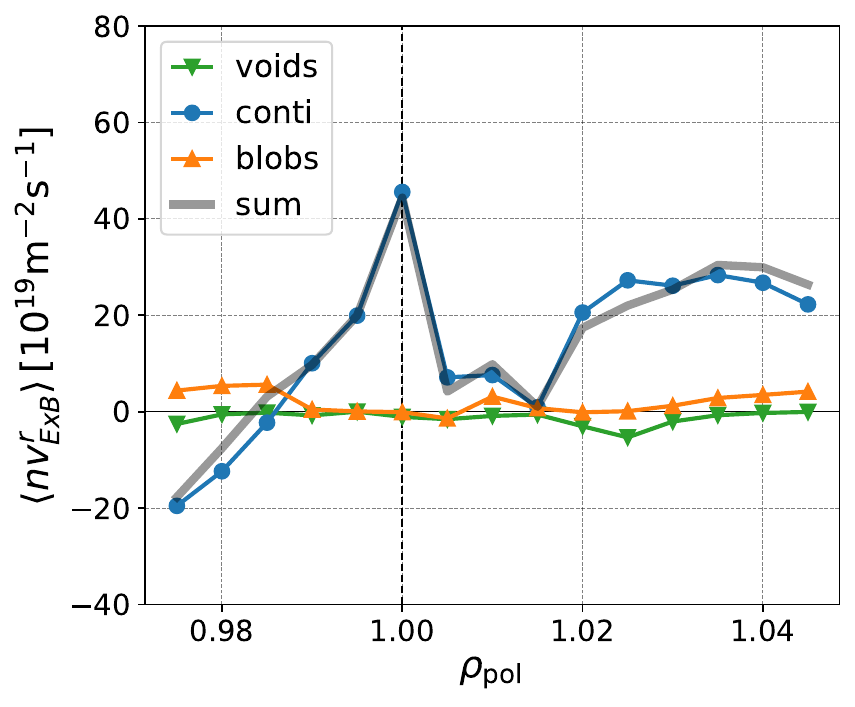}
    
\end{minipage}%
}
\subfloat[high-$\nsep$ \label{fig:qce:intermitt_exb_rho_h}]{%
\begin{minipage}[c]{0.46\textwidth}
\centering
    \includegraphics[width=\textwidth]{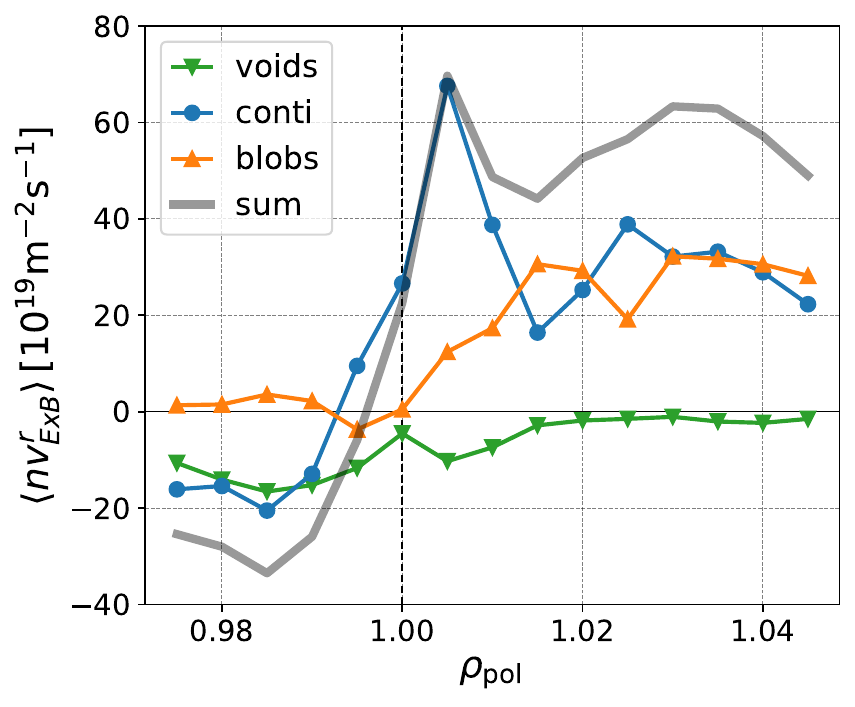}
    
\end{minipage}%
}
    \caption[Intermittency of SOL transport across radii]{Radial $E\times B$ particle flux decomposed according to Equations~\eqref{eqn:transport:void}--\eqref{eqn:transport:blob} into void, background, and blob contributions, together with their sum, for (a) the low-$\nsep$ and (b) the high-$\nsep$ case.}
    \label{fig:qce:intermitt_exb_rho}
\end{figure}

To obtain an additive decomposition, we categorize the instantaneous radial $E\times B$ flux according to the normalized density fluctuation $\tilde n/\sigma_n$. Specifically, we define three transport channels using indicator functions:
\begin{align}
    \Gamma_\mathrm{void}
    &=\left\langle nv_E^r\,\mathbf{I}_{\tilde n/\sigma_n<-2.5}\right\rangle_{t,\varphi},
    \label{eqn:transport:void}\\
    \Gamma_\mathrm{conti}
    &=\left\langle nv_E^r\,\mathbf{I}_{-2.5\leq \tilde n/\sigma_n\leq2.5}\right\rangle_{t,\varphi},
    \label{eqn:transport:conti}\\
    \Gamma_\mathrm{blob}
    &=\left\langle nv_E^r\,\mathbf{I}_{\tilde n/\sigma_n>2.5}\right\rangle_{t,\varphi}.
    \label{eqn:transport:blob}
\end{align}
Here, $\mathbf{I}$ is the indicator function. 
Each contribution includes both the event probability and its signed instantaneous flux, and the three components satisfy $\Gamma_\mathrm{sum}=\Gamma_\mathrm{void}+\Gamma_\mathrm{conti}+\Gamma_\mathrm{blob}$.
Figure~\ref{fig:qce:intermitt_exb_rho} shows the resulting components for $0.975\leq\rpol\leq1.045$. 
In the low-$\nsep$ case, non-intermittent fluctuations (blue markers) carry nearly all the flux. 
The flux peaks near the separatrix and decreases rapidly, reaching a minimum by $\rpol=1.005$, consistent with the experimental measurements in the EDA regime~\cite{grenfell2024multi}. 
The far-SOL rise of continuous-fluctuation-driven flux (blue markers) coincides with the emergence of the secondary separatrix at $\rpol=1.026$, which has been observed in both low- and high-$\nsep$ cases.
In the high-$\nsep$ case, the total flux rises and remains elevated across the SOL because the blob contribution becomes strongly positive. 
For $\rpol>1.01$, blobs carry approximately 50\% of the net $E\times B$ particle flux, while the continuous component remains comparable in magnitude and radial shape to that in the low-$\nsep$ state.
This blob-driven transport contrasts sharply with that in the low-$\nsep$ case.

\begin{figure}[!ht]
    \centering
    \includegraphics[width=0.5\linewidth]
    {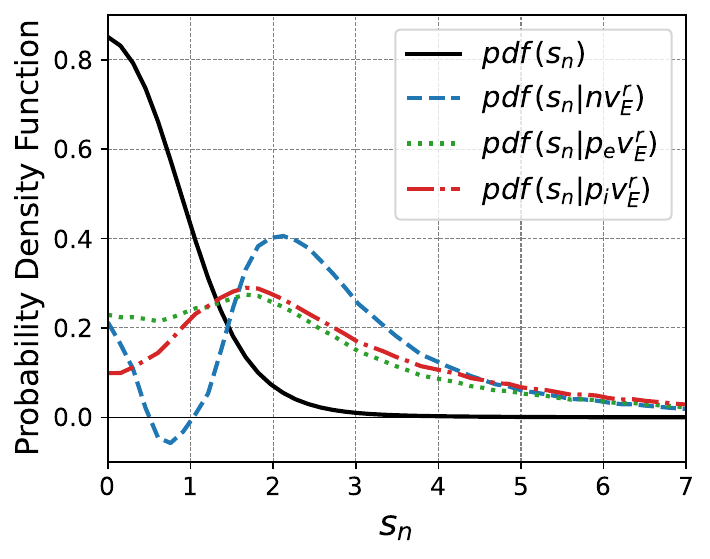}
    \caption[Intermittency probability and transport spectra in the far SOL]
    {Intermittency probability and transport spectra in the high-$\nsep$ case,
    evaluated over the low-field-side SOL region
    $1.03\leq\rpol\leq1.05$. The solid curve shows the probability
    density with respect to
    $s_n=|\tilde n/\sigma_n|$. The dashed, dotted, and dash--dotted
    curves show the corresponding flux-weighted distributions by the radial
    particle flux $nv_E^r$, electron heat flux $p_ev_E^r$, and ion heat
    flux $p_iv_E^r$, respectively. 
    All the probability density distributions are  normalized, so that their integrals are unity.}
    \label{fig:qce:flux_pdf_SOL_heat}
\end{figure}

We emphasize that the approximately 50\% contribution refers to the fraction of the radial flux carried by events satisfying the ${\tilde n}/{\sigma_n}>2.5$ criterion; it does not represent the fraction of particles contained within blobs. Its physical significance is that rare intermittent events make an order-unity contribution to the mean SOL transport, as demonstrated below.
We focus on the high-$\nsep$ case and examine the transport statistics in the far-SOL region $1.03\leq\rpol\leq1.05$, where blob-mediated transport is dominant. 
Frequent, low-amplitude density fluctuations produce large positive flux contributions for $0<\tilde n/\sigma_n<1$ and negative contributions for $-1<\tilde n/\sigma_n<0$. These opposing contributions largely cancel, leaving a much smaller net flux from the range $|\tilde n/\sigma_n|<1$.
To quantify the transport associated with fluctuations of different amplitudes, we introduce the absolute intermittency variable
\begin{equation}
    s_n=\left|{\tilde n}/{\sigma_n}\right|.
\end{equation}
Figure~\ref{fig:qce:flux_pdf_SOL_heat} compares the probability distribution of $s_n$ with the corresponding flux-weighted distributions. 
The solid curve represents the probability density function of $s_n$, such that its integral over a given interval gives the probability of observing fluctuations within that amplitude range.
The dashed curve shows the distribution weighted by the signed radial particle flux $nv_E^r$ (negative values for inward flux), and therefore describes how fluctuations of different amplitudes contribute to particle transport. 
Similarly, the dotted and dash--dotted curves show the distributions weighted by the electron and ion radial heat fluxes, $p_ev_E^r$ and $p_iv_E^r$, respectively.
No significant void population is present in this SOL region. The condition $s_n>2.5$ is therefore effectively equivalent to the blob criterion $\tilde n/\sigma_n>2.5$. Events satisfying this criterion account for only $1.8\%$ of all samples, as obtained by integrating the solid curve over $s_n \in[2.5,+\infty]$. 
In contrast, the corresponding area under the particle-flux spectrum shows that these events carry $56\%$ of the net radial particle flux. The numerical values are summarized in Table~\ref{table:qce:flux_pdf_1p03}.
The same blobs also make substantial contributions to radial heat transport, carrying $40\%$ of the electron heat flux and $47\%$ of the ion heat flux. Their fractional contributions to heat transport are somewhat smaller than their contribution to particle transport. 
This difference is also observed for ELM filaments in experiments \cite{boedo2005edge,yang2018combined,muller2011latest}, consistent with the stronger parallel energy losses along open field lines toward the divertor targets in the far SOL. 
If a less restrictive intermittency threshold of $s_n>1.5$ is adopted, the selected events account for $11\%$ of all samples but carry $91\%$ of the net particle flux, $65\%$ of the electron heat flux, and $74\%$ of the ion heat flux. These results demonstrate that particle and heat transport in the high-$\nsep$ SOL are both strongly dominated by the low-probability, large-amplitude tail of the fluctuation distribution.

\begin{table}[!ht]
    \centering
    \begin{tabular}{l|cc}
        \hline
        Quantity & $s_n>2.5$ & $s_n>1.5$ \\
        \hline
        Event probability
            & $1.8\%$ & $11\%$ \\
        Contribution to $nv_E^r$
            & $56\%$ & $91\%$ \\
        Contribution to $p_ev_E^r$
            & $40\%$ & $65\%$ \\
        Contribution to $p_iv_E^r$
            & $47\%$ & $74\%$ \\
        \hline
    \end{tabular}
    \caption[Event probabilities and transport fractions]
    {Event probabilities obtained using the intermittency thresholds
    $s_n>2.5$ and $s_n>1.5$, together with the corresponding fractions
    of the net radial particle flux $nv_E^r$, electron heat flux
    $p_ev_E^r$, and ion heat flux $p_iv_E^r$. The values are obtained by
    integrating the corresponding distributions in
    Figure~\ref{fig:qce:flux_pdf_SOL_heat} beyond each threshold.}
    \label{table:qce:flux_pdf_1p03}
\end{table}
\section{Mechanism of QCM-driven blob launching}
Tokamak SOL blobs generally emerge through the nonlinear evolution of edge turbulence, but the detailed formation mechanism depends on the confinement regime. Blob generation is commonly associated with interchange dynamics and sheared ${E}\times{B}$ flows~\cite{d2011convective,Furno2008}. 
As shown above, both cases contain a KBM-driven QCM, whereas only the high-$\nsep$ case contains the RXM and launches blobs. 
This comparison motivates an examination of the two candidate ingredients in the launching process: ${E}\times{B}$ flow shear and interchange-like dynamics.

\subsection{Role of \texorpdfstring{$E\times B$}{E x B} shear}
\label{sec:qce:er}

\begin{figure}
    \centering
    \includegraphics[width=0.5\linewidth]{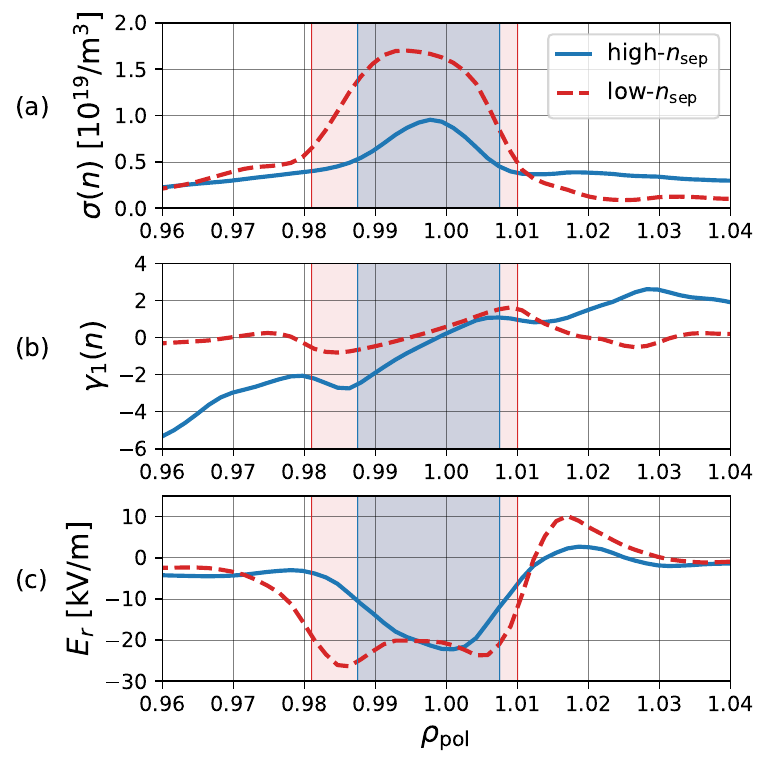}
    \caption[Radial electric field and density-fluctuation statistics]{Outboard-midplane profiles of (a) the density-fluctuation standard deviation $\sigma_n$, (b) the density-fluctuation skewness $\gamma_1(n)$, and (c) the radial electric field $E_r$. The vertical lines bound the region between the extrema of the $E_r$ shear.}
    \label{fig:qce:Er}
\end{figure}

Figure~\ref{fig:qce:Er} shows the outboard-midplane profiles of the density-fluctuation standard deviation and skewness, together with the radial electric field, on the same $\rpol$ coordinate. 
In both cases, the strong QCM fluctuations near the separatrix, quantified by $\sigma_n$, are consistently encompassed by a pronounced $E_r$ shear layer. This region is indicated by the shaded area, which is slightly wider in the low-$\nsep$ case.
The large positive skewness in the SOL of the high-$\nsep$ case is consistent with the blobs identified in Figure~\ref{fig:qce:pdf_rho_list}, whereas the skewness remains close to zero in the SOL of the low-$\nsep$ case. 
Within the shaded region, the QCM skewness changes from negative to positive in both cases. 
A radial skewness reversal develops within both QCM envelopes, but only the high-$\nsep$ case has clear blobs outside the QCM range.

The $E\times B$ shear alone cannot explain the blob generation from the QCM in the SOL. 
In fact, the low-$\nsep$ case exhibits an even stronger $E_r$ shear at the outer edge of the QCM fluctuation region, while no blobs are detected in the SOL. 
The QCM and the $E_r$ profile coevolve and regulate one another through the radial force-balance relation $E_r = (\partial_r p_i)/(en) + u_{\varphi} B_{\theta} - u_{\theta} B_{\varphi}$. 
The intrinsic flow and magnetic nonlinearity associated with the QCM can modify the poloidal flow $u_{\theta}$, while QCM-driven transport can reshape the ion pressure gradient $\partial_r p_i$. In turn, the $E_r$ shear layer limits the radial extent of the QCM. 
The QCM and the $E_r$ shear layer should therefore be viewed as a coupled dynamical state, rather than treating the QCM as a mode evolving on a fixed $E_r$ background.
The role of the $E_r$ shear is primarily to regulate the radial location and width of the QCM. In the saturated phase, the QCM remains confined within the $E_r$ well. The strongest $E\times B$ shear is located outside the region of intense QCM fluctuations and therefore cannot cut off QCM fragments or account for the continuous ejection of blobs. 
The $E\times B$ shear can advect existing blobs poloidally away from the QCM as shown in Ref.\cite{zhang2026turbulent}. However, it cannot explain how the blobs are generated from the QCM in the first place.

\subsection{Interchange conversion}\label{sec:qce:interchange}

\begin{figure}
    \centering
\subfloat[low-$\nsep$\label{fig:qce:blob_generate_pot_EDA}]{%
\begin{minipage}[c]{0.478\textwidth}
\centering
    \includegraphics[width=\textwidth]{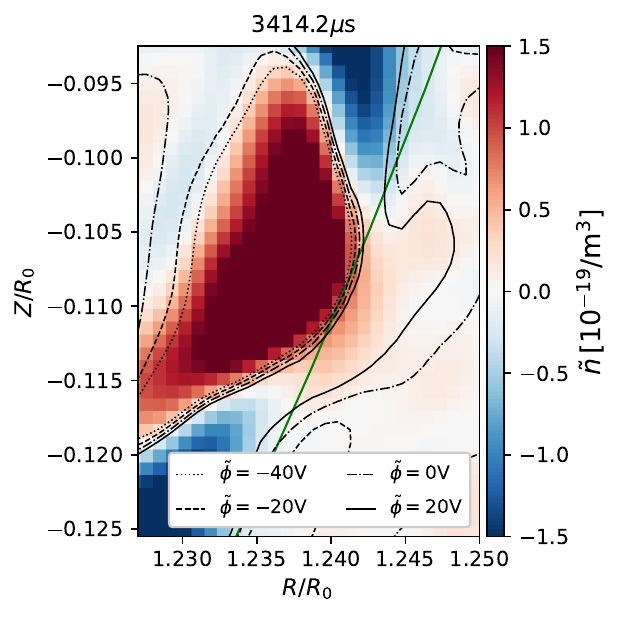}

\end{minipage}%
}
\subfloat[high-$\nsep$\label{fig:qce:blob_generate_pot_QCE}]{%
\begin{minipage}[c]{0.478\textwidth}
\centering
   \includegraphics[width=\textwidth]{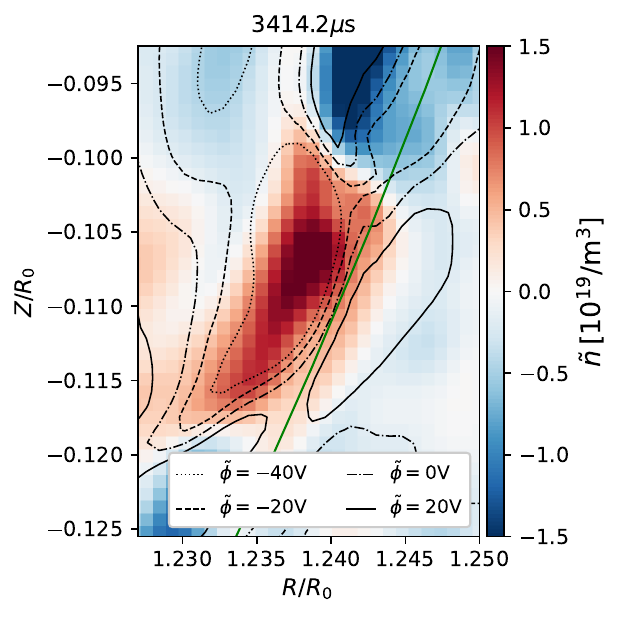}

\end{minipage}%
}
\caption[Potential structure during blob launching]{Density fluctuations $\tilde n$ (color) and electrostatic-potential contours $\tilde\phi$ in (a) the low-$\nsep$ case, where no blob is launched, and (b) the high-$\nsep$ case during the birth of a blob. 
Dashed, dash--dotted, and solid contours denote $\tilde\phi<0$, $\tilde\phi=0$, and $\tilde\phi>0$, respectively; the contour spacing is $20\,\si{V}$.}\label{fig:qce:blob_generate_pot}
\end{figure}

Our simulations instead indicate that interchange-like dynamics play a crucial role in blob formation.
This is evident in the two-dimensional correlation between $\tilde{n}$ and $\tilde{\phi}$ during a blob-launching event, as shown in Figure~\ref{fig:qce:blob_generate_pot_QCE}.  
The main body of the positive $\tilde{n}$ fluctuation associated with the QCM lies beneath the negative $\tilde{\phi}$ contours (dashed lines), consistent with the anti-phase correlation in Figure~\ref{fig:qce:fourier_shift_QCE}.  
At the outermost radial layer of the QCM, $\tilde{\phi}$ develops a butterfly-shaped pattern, indicating a sudden change in phase correlation and $E\times B$ advection along the open, outward-directed $\tilde{\phi}=0$ contour (dash--dotted line).
This structure suggests a transition from kinetic-shear-Alfv\'en-like oscillations to interchange-like dynamics.
This local interchange structure facilitates the radial detachment of filaments from the QCM and their subsequent propagation into the SOL.
In contrast, the low-$\nsep$ case in Figure~\ref{fig:qce:blob_generate_pot_EDA} shows that most positive $\tilde{n}$ fluctuations remain confined within the $\tilde{\phi}\leq0$ region.  
The absence of an interchange-like potential structure suggests that the fluctuating $E\times B$ drift primarily rotates the QCM packet along closed $\tilde{\phi}$ contours without expelling fragments from the QCM region.

\begin{figure}
    \centering
    
\subfloat[\label{fig:qce:shift_1p004_l}]{%
    
\begin{minipage}[b]{0.45\textwidth}
    
\centering
    \includegraphics[width=\textwidth]{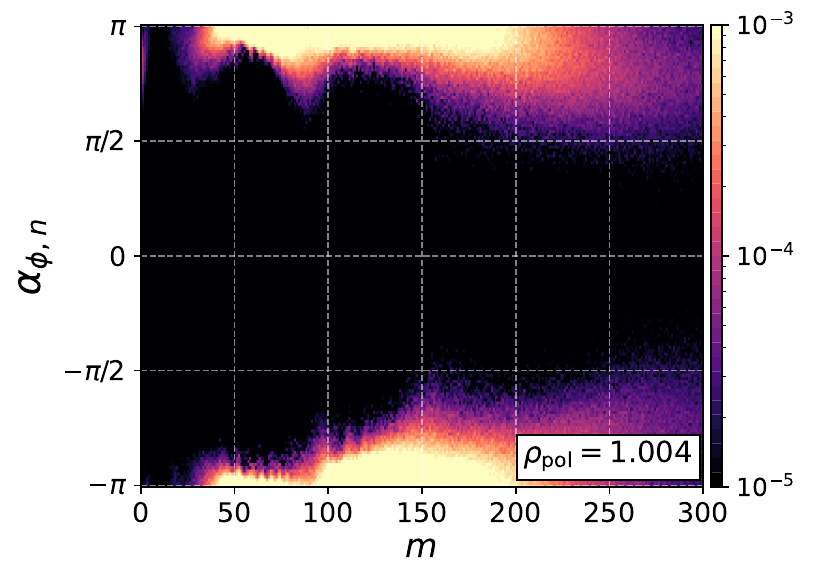}

\end{minipage}%
    
}
\subfloat[\label{fig:qce:shift_1p004_h}]{%
\begin{minipage}[b]{0.45\textwidth}
\centering
    \includegraphics[width=\textwidth]{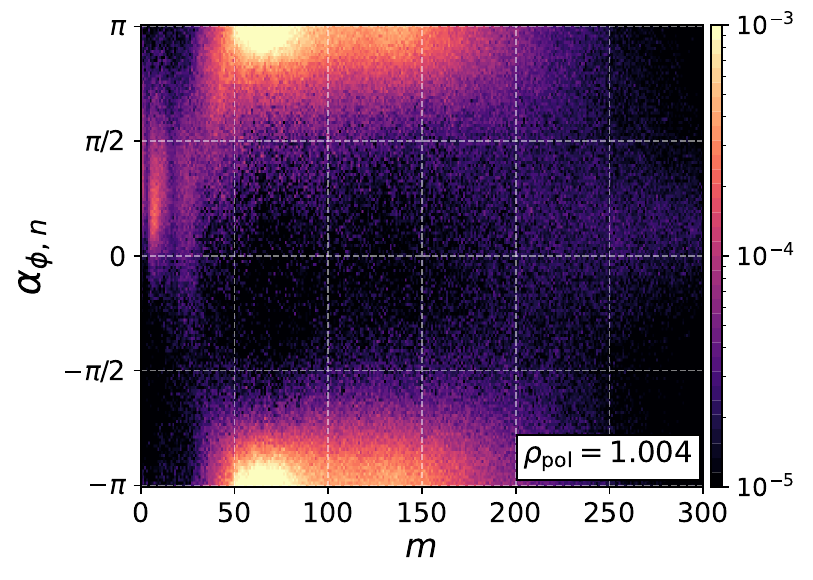}
    
\end{minipage}%
}

\subfloat[\label{fig:qce:shift_1p008_l}]{%

\begin{minipage}[b]{0.45\textwidth}

\centering
    \includegraphics[width=\textwidth]{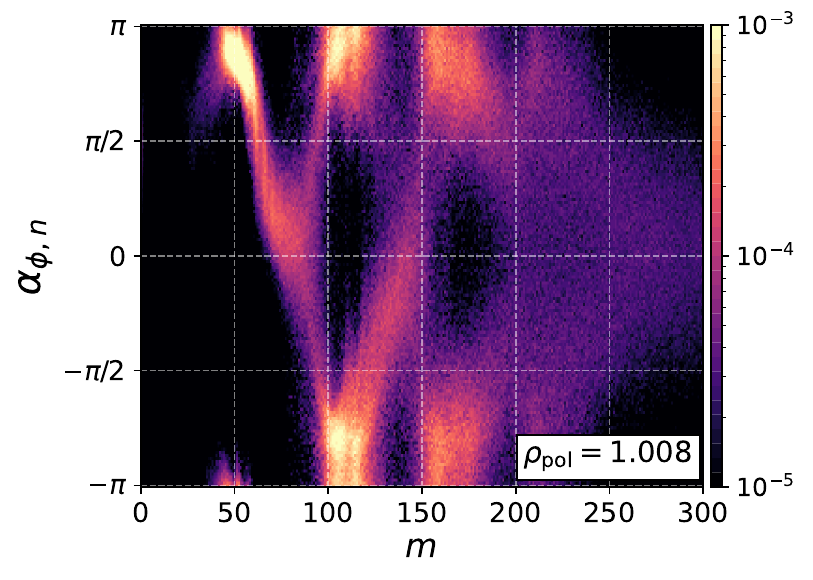}

\end{minipage}%

}
\subfloat[\label{fig:qce:shift_1p008_h}]{%
\begin{minipage}[b]{0.45\textwidth}
\centering
    \includegraphics[width=\textwidth]{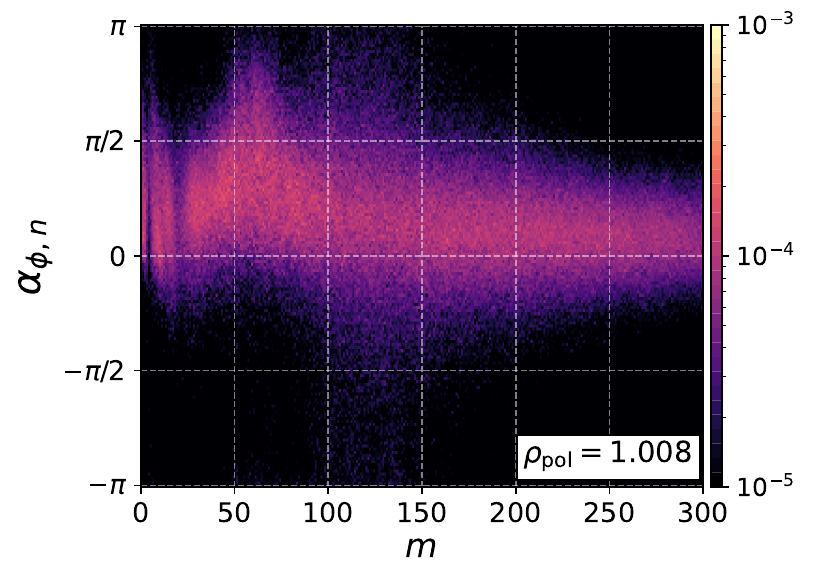}
    
\end{minipage}%
}

\subfloat[low-$\nsep$\label{fig:qce:shift_1p012_l}]{%

\begin{minipage}[b]{0.45\textwidth}

\centering
    \includegraphics[width=\textwidth]{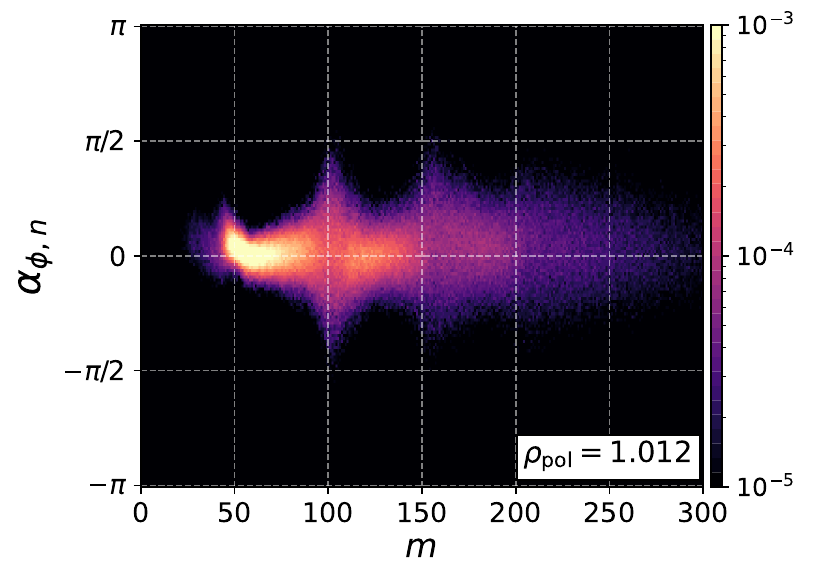}

\end{minipage}%

}
\subfloat[high-$\nsep$\label{fig:qce:shift_1p012_h}]{%
\begin{minipage}[b]{0.45\textwidth}
\centering
    \includegraphics[width=\textwidth]{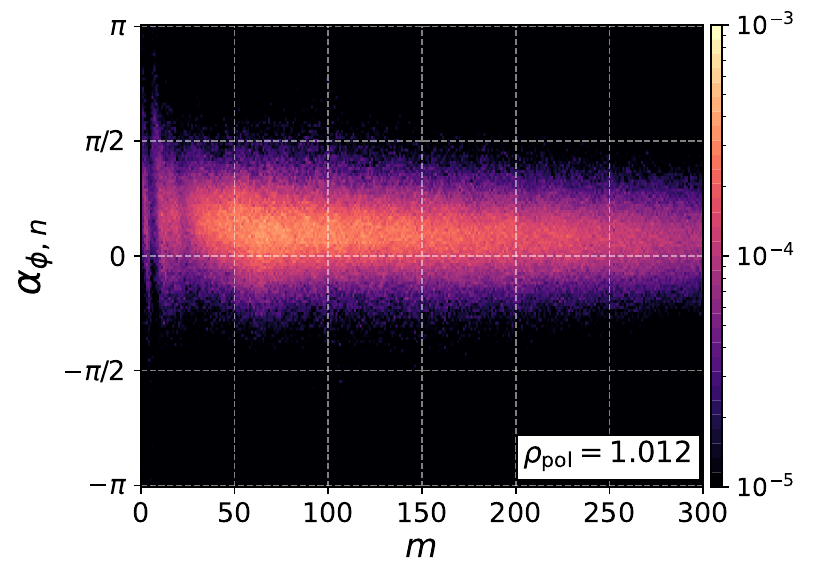}
    
\end{minipage}%
}

\caption[Phase shift $\alpha_{\phi,n}$ of QCM across radii]{Evolution of the phase shift $\alpha_{\phi,n}$ with increasing $\rpol$ in the SOL. The right column illustrates the high-$\nsep$ case, which includes a subdominant resistive mode. The left column provides a comparison with the low-$\nsep$ case, where this mode is absent.}
\label{fig:qce:shift_sol}
\end{figure}

The Fourier spectra of the phase shift in Figure~\ref{fig:qce:shift_sol} show how the conversion develops. Because a unique binormal wavenumber is unavailable on open field lines, the horizontal axis is the poloidal mode number $m$. At $\rpol=1.004$, both cases retain the QCM peak near $m=50$ and $\alpha_{\phi,n}=\pi$. Only the high-$\nsep$ case contains an additional low-$m$ RXM component. Its cross-phase changes from approximately $-\pi/4$ inside the separatrix to $+\pi/4$ in the SOL.
With increasing radius, the high-$\nsep$ QCM cross-phase changes continuously from $\pi$ at $\rpol=1.004$ to approximately $\pi/2$ at $\rpol=1.008$. By $\rpol=1.012$, the QCM peak has faded into a spectrum centered near $\alpha_{\phi,n}=\pi/4$. The electron- and ion-temperature cross-phases evolve similarly, so the blobs contain in-phase positive $n$, $T_e$, and $T_i$ perturbations, consistent with experimental blob measurements~\cite{boedo2003transport,Endler2005}.
In the low-$\nsep$ case, which lacks the RXM, the cross-phase changes abruptly from $\pi$ to approximately zero and does not form a robust $\pi/2$ band at the QCM scale. The comparison supports a mechanism in which KBM--RXM coupling drives an interchange-like phase conversion that launches blobs.

\subsection{Dynamical transition and parameter dependence}
\label{sec:qce:transition}

\begin{figure}
    \centering
    \includegraphics[width=0.87\linewidth]{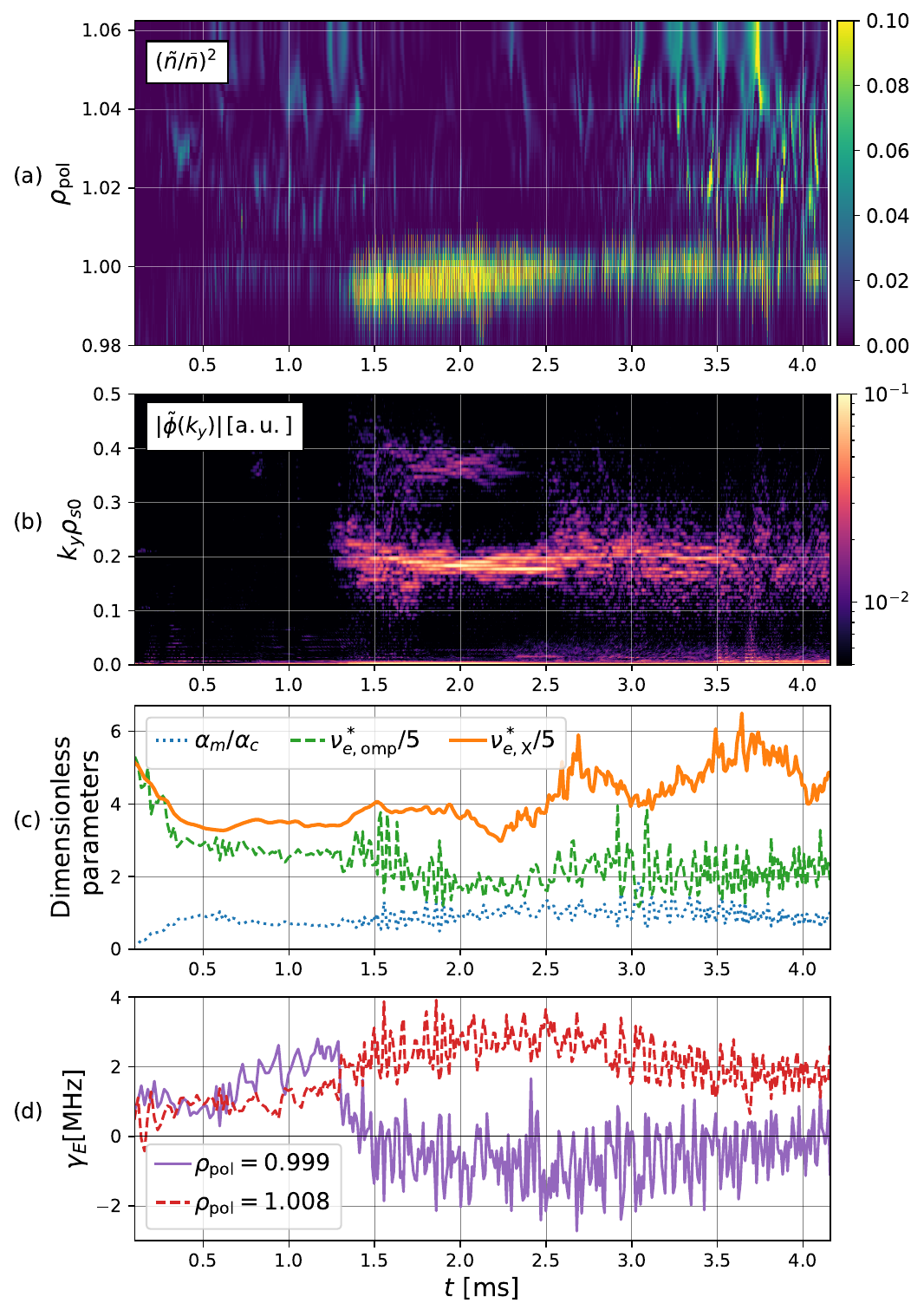}
    \caption[Transition from a coherent QCM to blob-producing turbulence]{Time evolution of the high-$\nsep$ case: (a) relative outboard-midplane density-fluctuation amplitude versus radius; (b) Fourier amplitude of $\tilde\phi$ at $\rpol=0.999$; (c) dimensionless parameters measuring the MHD ballooning drive $\alpha_m/\alpha_c$ and electron collisionality at the OMP, $\nu_{e,\mathrm{omp}}^*$, and near the X-point, $\nu_{e,\mathrm{X}}^*$ (Both collisionality curves are divided by 5 for display.); and (d) outboard-midplane $E\times B$ shear rate at $\rpol=0.999$ (QCM center) and $\rpol=1.008$ (blob-launching surface).}
    \label{fig:qce:mode_overview_h}
\end{figure}

The above interpretation is further supported by the dynamical transition observed in the high-$\nsep$ simulation. 
Figure~\ref{fig:qce:mode_overview_h}(a) shows the time evolution of the density fluctuations at the outboard midplane. 
On the same time axis, Figure~\ref{fig:qce:mode_overview_h}(b) shows the Fourier spectrum at the flux surface $\rpol=0.999$. 
The QCM emerges around $t=1.5\,\si{ms}$ in the vicinity of the separatrix, while blobs in the SOL first appear at approximately $t=3\,\si{ms}$. 
These two events mark two important transitions during the simulation.
During the interval $[1.5, 2.5]\,\si{ms}$, the QCM does not produce blobs. As shown in Figure~\ref{fig:qce:mode_overview_h}(b), the KBM is present at $k_y\rho_s=0.2$ together with higher harmonics, whereas the RXM is absent. 
This stage is consistent with the low-$\nsep$ case. 
In the later stage, after $t=2.5\,\si{ms}$, the RXM emerges at low $k_y$ and begins to interact with the KBM spectrum. 
The KBM becomes more broadband around $k_y\rho_s=0.2$, while its higher harmonics disappear. 
Blobs are detected subsequently, as shown in Figure~\ref{fig:qce:mode_overview_h}(a).

To facilitate the interpretation, Figure~\ref{fig:qce:mode_overview_h}(c) shows three key dimensionless parameters relevant to the microinstabilities. 
The MHD ballooning parameter, normalized to the ideal ballooning threshold, $\alpha_m/\alpha_c$, provides a measure of the proximity to the KBM threshold. 
Here, $\alpha_c$ is constant and equal to $2.7$ for this equilibrium. 
The electron collisionality at the outboard midplane, $\nu_{e,\mathrm{omp}}^*$ (evaluating Eq.~\ref{eqn:nu} with $n_{\mathrm{sep}}$ and $T_{e,\mathrm{sep}}$), is relevant to the destabilization of resistive ballooning modes and collisional drift waves. 
We note that the turbulence control parameter $\alpha_t$ \cite{eich2020turbulence} is proportional to $\nu_{e,\mathrm{omp}}^*$ in this simulation, as the magnetic equilibrium is fixed. 
The near-X-point collisionality, $\nu_{e,\mathrm{X}}^*$, is evaluated at the orthogonal projection of the X-point onto the flux surface $\rpol=0.999$ and serves as a local control parameter for the RXM.
(Quantities evaluated at the exact X-point are less representative because the vanishing poloidal magnetic field leads to a formally divergent parallel connection length and hence a decoupling from the upstream plasma.)
Figure~\ref{fig:qce:mode_overview_h}(d) shows the $E\times B$ shear rate, $\gamma_E$, at two radial locations: $\rpol=0.999$, corresponding approximately to the center of the QCM, and $\rpol=1.008$ at the outermost layer of the QCM. 
During the simulation, the magnetic shear and plasma shaping remain fixed by the equilibrium, while those dimensionless parameters and the $E\times B$ shear rates evolve as the plasma profiles change.

The emergence of the RXM and the subsequent appearance of blobs coincide with an increase in $\nu_{e,\mathrm{X}}^*$ to approximately $25\sim30$ around $t=2.5\,\si{ms}$. 
(We note that the increase in $\nu_{e,\mathrm{X}}^*$ is associated with the ramp-up of the global ionization rate, which more than doubles from $S_{\mathrm{iz}}=5.6\cdot10^{22}/\si{s}$ at $t=1.5\,\si{ms}$ to $S_{\mathrm{iz}}=1.3\cdot10^{23}/\si{s}$ after $t=3\,\si{ms}$.)
By contrast, the evolution of the upstream collisionality, $\nu_{e,\mathrm{omp}}^*$, cannot explain the transition to the blob-launching stage. 
(Consistently, the low-$\nsep$ case has $\nu_{e,\mathrm{X}}^*=10$, which remains below the corresponding value in the high-$\nsep$ case.)
Moreover, a high $\gamma_E$ in the SOL is not required for blob generation; it actually decreases with the onset of blobs. 
This supports our interpretation that blob launching results from interchange-like dynamics promoted by the RXM, rather than from $E\times B$ shear cutting fragments off the QCM.

The main drive for the KBM, characterized by $\alpha_m$, approaches the ideal ballooning threshold $\alpha_m/\alpha_c\sim 1$ as early as $t=0.5\,\si{ms}$, well before the QCM is observed at $t=1.5\,\si{ms}$. 
However, during the interval $[0.5,1.5]\,\si{ms}$, the strong $E\times B$ shear at $\rpol=0.999$ suppresses the growth of QCM. 
The onset of the QCM at $t=1.5\,\si{ms}$ coincides with a significant reduction of the shear rate to approximately $\gamma_E=0$ at $\rpol=0.999$. 
This result emphasizes the role of nonlinear physics and turbulence self-organization in the formation of the QCM, as discussed in Ref.\cite{zhang2026turbulent}.
The upstream electron collisionality, $\nu_{e,\mathrm{omp}}^*$, shows no corresponding increase at the QCM onset. 
This observation provides further evidence against interpreting the QCM as a collisionality-destabilized mode and is instead consistent with its identification as a KBM.

\section{Conclusions}\label{sec:conclusion}

We have investigated the physical origin of the quasicoherent mode (QCM) in quasicontinuous exhaust (QCE) ELM-free H-mode tokamak plasmas, as well as the mechanism by which QCM perturbations develop into SOL blobs. Global, nonlinear, full-$f$, two-fluid electromagnetic simulations were compared for two edge states with different separatrix densities in the same magnetic geometry. 
The states were obtained using different treatments of neutral gas boundary conditions. 
They constitute a comparison between two self-consistent edge conditions rather than a scan in $\nsep$ alone.

In both states, the QCM drives a radial oscillation of the pedestal foot across the separatrix, a behavior distinct from conventional L-mode and inter-ELM H-mode turbulence. 
Its field alignment, twisting parity, electromagnetic character, characteristic scale $k_y\rho_s\approx0.2$, ion-diamagnetic propagation, plasma-frame frequency, and potential--density cross-phase near $\pi$ identify a kinetic ballooning mode as the main drive. 
The stronger coherence and higher harmonics in the low-$\nsep$ case resemble the experimental characteristics of the EDA regime in ASDEX Upgrade, while the high-$\nsep$ case has more broadband QCM and was previously validated against the QCE experiment \cite{zhang2026turbulent}.

The presence of a KBM-driven QCM is not, however, sufficient to launch blobs. Although the fluctuation skewness changes from negative to positive across the QCM envelope in both cases, the SOL blobs appear only in the high-$\nsep$ state. 
These blobs originate from positive-density perturbations that detach from the outer edge of the QCM near $\rpol\approx1.01$ and subsequently propagate into the SOL. 
For $\rpol>1.01$, those blob events with an occurrence probability below $2\%$ account for approximately half of the net radial $E\times B$ particle and heat fluxes and therefore make an order-unity contribution to the mean SOL transport.
The $E\times B$ shear regulates the radial position and width of the QCM and advects existing blobs poloidally, but it does not explain their formation. 
Instead, blob ejection is accompanied by a change in the potential--density cross-phase from the KBM-like value near $\pi$ toward the interchange value near $\pi/2$. 
This conversion is associated with an additional resistive X-point mode (RXM), which originates from the X-point and appears at lower wavenumbers $k_y$ than the QCM. 

The dynamical transition of the high-$\nsep$ case provides further support for those interpretations.
The QCM emerges after the MHD ballooning parameter $\alpha_m$ approaches the ideal ballooning threshold $\alpha_c$ and $E\times B$ shear weakens.
The RXM and blobs appear only after the near-X-point collisionality rises. 
Neither of the corresponding transitions is associated with a rising upstream collisionality.

Taken together, these results suggest that the QCE regime should be understood not as the manifestation of a single instability of QCM, but as a coupled edge state. The KBM-driven QCM provides continuous pedestal regulation, while the RXM enables an interchange-like conversion that opens a pathway from coherent edge fluctuations to intermittent SOL exhaust. This separation of roles implies that predictive access to reactor-relevant QCE operation requires accounting not only for the dominant edge instability, but also for its spatially structured coupling to X-point physics.

\section*{Conflict of Interest}
The authors have no conflicts to disclose.








 
\section*{Acknowledgements}
This work received funding from the German Federal Ministry of Research, Technology and Space (BMFTR) within the funding program Fusion 2040 (Reference No. 13F1046). This work was carried out within the framework of the EUROfusion Consortium, funded by the European Union via the Euratom Research and Training Programme (Grant Agreement No. 101052200—EUROfusion). The views and opinions expressed are however those of the author(s) only and do not necessarily reflect those of the European Union or the European Commission. Neither the European Union nor the European Commission can be held responsible for them. The simulations were performed on the national supercomputer HPE Apollo Hawk at the High Performance Computing Center Stuttgart (HLRS) under Grant No. GRILLIX/44281.

\section*{Appendix}\label{appe}
\begin{figure}
    \centering
    \includegraphics[width=0.5\linewidth]{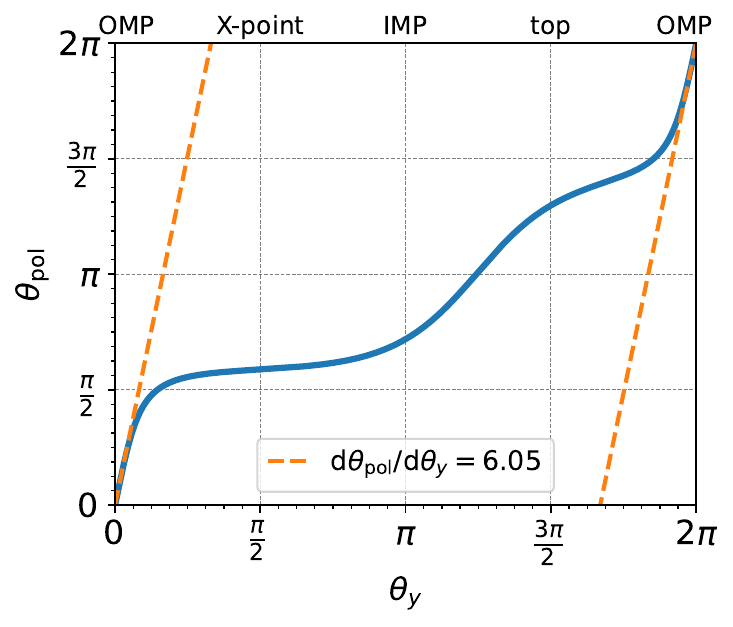}
    \caption[Coordinate mapping: $\theta_y$ to $\theta_\mathrm{pol}$]{Variation of the conventional poloidal angle $\theta_\mathrm{pol}$ as a function of the symmetry coordinate $\theta_y$ at the flux surface $\rho_\mathrm{pol} = 0.999$. Key poloidal locations are marked: outboard midplane (OMP), X-point, inboard midplane (IMP), and top of the tokamak.}
    \label{fig:qce:theta_y_pol}
\end{figure}

Figure~\ref{fig:qce:theta_y_pol} shows the mapping relationship between the symmetry coordinate $\theta_y$ and the conventional poloidal angle $\theta_\mathrm{pol}$ at the flux surface $\rpol=0.999$.

\bibliographystyle{unsrt}
\bibliography{references}

\section*{ASDEX Upgrade Collaboration Author List}
\begingroup\small\noindent
T. Pütterich\textsuperscript{1},
F. Albrecht\textsuperscript{1,7},
E. Alessi\textsuperscript{12},
S. An\textsuperscript{1},
C. Angioni\textsuperscript{1},

F. Antlitz\textsuperscript{1},
N. Arden\textsuperscript{1},
V. Artigues\textsuperscript{1},
M. Astrain\textsuperscript{1},
M. Balden\textsuperscript{1}
% ...
\par\medskip
\noindent\textbf{Affiliations}\par\smallskip
\noindent\textsuperscript{1} Max-Planck-Institut für Plasmaphysik, Garching, Germany\par
\noindent\textsuperscript{2} Max-Planck-Institut für Plasmaphysik, Greifswald, Germany\par
\noindent\textsuperscript{3} Instituto de Plasmas e Fusão Nuclear, Instituto Superior Técnico, Universidade de Lisboa, Portugal\par
\noindent\textsuperscript{4} Department of Applied Physics, Aalto University, Espoo, Finland\par
\noindent\textsuperscript{5} Eindhoven, University of Technology, Eindhoven, Netherlands\par
\noindent\textsuperscript{6} CCFE, Culham Science Centre, Abingdon, United Kingdom of Great Britain and Northern Ireland\par
\noindent\textsuperscript{7} Physics Department, Technical University of Munich, TUM School of Natural Sciences, 85748 Garching, Germany\par
\noindent\textsuperscript{8} Consorzio RFX, Padova, Italy\par
\noindent\textsuperscript{9} CEA/IRFM, Saint Paul Lez Durance, France\par
\noindent\textsuperscript{10} Forschungszentrum Jülich, Germany\par
\noindent\textsuperscript{11} Institut Jean Lamour, Université de Lorraine, Nancy, France\par
\noindent\textsuperscript{12} ENEA, IFP-CNR, Milan, Italy\par
\noindent\textsuperscript{13} Ecole Polytechnique Fédérale de Lausanne (EPFL), Swiss Plasma Center (SPC), Lausanne, Switzerland\par
\noindent\textsuperscript{14} ÖAW, IAP, University of Innsbruck, Innsbruck, Austria\par
\noindent\textsuperscript{15} Centre for Energy Research, Budapest, Hungary\par
\noindent\textsuperscript{16} Institute of Plasma Physics and Laser Microfusion, Warsaw, Poland\par
\noindent\textsuperscript{17} Universität Greifswald, Germany\par
\noindent\textsuperscript{18} KTH Royal Institute of Technology, Stockholm, Sweden\par
\noindent\textsuperscript{19} General Atomics, San Diego, CA, United States of America\par
\noindent\textsuperscript{20} Universidad de Sevilla, Spain\par
\noindent\textsuperscript{21} VTT Technical Research Centre of Finland, VTT, Espoo, Finland\par
\noindent\textsuperscript{22} Budapest University of Technology and Economics, Budapest, Hungary\par
\noindent\textsuperscript{23} Laboratoire de Physique des Plasmas, Ecole Polytechnique, Palaiseau, France\par
\noindent\textsuperscript{24} IGVP Universität Stuttgart, Germany\par
\noindent\textsuperscript{25} Department of Physics, Technical University of Denmark, Kgs. Lyngby, Denmark\par
\noindent\textsuperscript{26} Department of Industrial Engineering, University of Rome, Italy\par
\noindent\textsuperscript{27} Ghent University, Ghent, Belgium\par
\noindent\textsuperscript{28} ERM/KMS, Brussels, Belgium\par
\noindent\textsuperscript{29} ÖAW, IAP, Vienna University of Technology, Austria\par
\noindent\textsuperscript{30} York Plasma Institute, University of York, United Kingdom of Great Britain and Northern Ireland\par
\noindent\textsuperscript{31} Electrical and Computer Engineering, University of California, Davis, CA, United States of America\par
\noindent\textsuperscript{32} Princeton Plasma Physics Laboratory, Princeton, NJ, United States of America\par
\noindent\textsuperscript{33} Barcelona Supercomputing Center, Barcelona, Spain\par
\noindent\textsuperscript{34} ENEA Consorzio CREATE, Naples, Italy\par
\noindent\textsuperscript{35} School of Physics, University College Cork, Cork, Ireland\par
\noindent\textsuperscript{36} Department of Materials Engineering, Technical University of Munich, TUM School of Engineering and Design, 85748 Garching, Germany\par
\noindent\textsuperscript{37} ISC-CNR, Politecnico di Torino, Italy\par
\noindent\textsuperscript{38} Department of Electrical and Electronic Engineering, University of Cagliari, Italy\par
\noindent\textsuperscript{39} Chinese Academy of Sciences, Hefei, China\par
\noindent\textsuperscript{40} Ioffe Institute, St. Petersburg, Russian Federation\par
\noindent\textsuperscript{41} MIT Plasma Science and Fusion Center, Cambridge, MA, United States of America\par
\noindent\textsuperscript{42} De Vinci Higher Education, De Vinci Research Center, 92 916 Paris, France\par
\noindent\textsuperscript{43} ITER Organization, Saint-Paul-lez-Durance, France\par
\noindent\textsuperscript{44} ENEA, University of Milano-Bicocca, Milano, Italy\par
\noindent\textsuperscript{45} Institute of Plasma Physics of the CAS, Praha, Czech Republic\par
\noindent\textsuperscript{46} University of Wisconsin, Madison, WI, United States of America\par
\noindent\textsuperscript{47} Department of Physics, Chalmers University of Technology, Gothenburg, Sweden\par
\noindent\textsuperscript{48} Laboratorio Nacional de Fusión, CIEMAT, Madrid, Spain\par
\noindent\textsuperscript{49} Department of Physics, Durham University, United Kingdom of Great Britain and Northern Ireland\par
\noindent\textsuperscript{50} ENEA, Centro Ricerche Frascati, Italy\par
\noindent\textsuperscript{51} Aix-Marseille University, CNRS, Marseille, France\par
\noindent\textsuperscript{52} Institute of Plasma Physics, National Science Center Kharkov Institute of Physics and Technology, Krakov, Ukraine\par
\noindent\textsuperscript{53} ÖAW, Graz University of Technology, Graz, Austria\par
\noindent\textsuperscript{54} Columbia University, New York City, NY, United States of America\par
\noindent\textsuperscript{55} CFS, Fort Devens, MA, United States of America\par
\medskip\noindent\footnotesize Source: T. Pütterich et al., ``Overview of the ASDEX Upgrade results, 	extit{Nuclear Fusion} 	extbf{66} (2026) 116002, DOI: 10.1088/1741-4326/ae61c8. Licensed under CC BY 4.0.
\endgroup

\end{document}